%% file: main.tex
\documentclass[prd,amsmath,aps,floats,amssymb, floatfix,
  superscriptaddress,nofootinbib,twocolumn,preprintnumbers]{revtex4-1}
\usepackage{amsmath,amssymb,natbib,latexsym,times}

\DeclareMathOperator\arctanh{arctanh}
\usepackage[switch,columnwise]{lineno}

\usepackage[T1]{fontenc}
\usepackage{aecompl}
\newcommand{\mr}{\mathrm}
\newcommand{\sfig}[2]{
\includegraphics[width=#2]{#1}
        }
\newcommand{\Sfig}[2]{
    \begin{figure}[thbp]
    \sfig{#1.pdf}{\columnwidth}
    \caption{{\small #2}}
    \label{fig:#1}
    \end{figure}
}
\newcommand{\Swide}[2]{
\begin{figure*}[thbp]
 \sfig{#1.pdf}{.8\textwidth}
  \caption{{\small #2}}
   \label{fig:#1}
   \end{figure*}
}
\newcommand{\Sswide}[2]{
\begin{figure*}[thbp]
 \sfig{#1.pdf}{.7\textwidth}
  \caption{{\small #2}}
   \label{fig:#1}
   \end{figure*}
}
\newcommand{\Svwide}[2]{
\begin{figure*}[thbp]
 \sfig{#1.pdf}{\textwidth}
  \caption{{\small #2}}
   \label{fig:#1}
   \end{figure*}
}

\newcommand{\rf}[1]{\ref{fig:#1}}
\newcommand{\ec}[1]{Eq.~(\ref{eq:#1})}
\newcommand{\ecalt}[1]{Eq.~\ref{eq:#1}}
\newcommand{\Ec}[1]{(\ref{eq:#1})}
\newcommand{\eeec}[3]{Eqs.~(\ref{eq:#1}, \ref{eq:#2}, \ref{eq:#3})}
\newcommand{\eql}[1]{\label{eq:#1}}
\def\vs{\nonumber\\}

\newcommand{\aj}{A.J.}
\newcommand{\mnras}{MNRAS}

\newcommand{\jcap}{JCAP}

\usepackage{verbatim}
\usepackage[usenames,dvipsnames]{xcolor}
\usepackage{tabulary}
\usepackage{tabularx}
\usepackage{todonotes}
\usepackage{hyperref}
\numberwithin{equation}{section}

\newcommand\salpha{{\eta_{\rm IA}}}

\newcommand{\photoz}{photo-$z$}

\newcommand\lcdm{$\Lambda$CDM}
\newcommand\wcdm{$w$CDM}

\newcommand{\imshape}{{\textsc{im3shape}}}

\newcommand\metacal{{\textsc{metacalibration}}} %{{\tt metacal}}
\newcommand\im{{\textsc{im3shape}}} %{{\tt im3shape}}

\newcommand\be{\begin{equation}}
\newcommand\ee{\end{equation}}
\def\bea{\begin{eqnarray}}
\def\eea{\end{eqnarray}}
\newcounter{syscounter}  % So we can continue an enumeration of robustness checks

\allowdisplaybreaks

\begin{document}
%\linenumbers

\title{Dark Energy Survey Year 1 Results:\\ Cosmological Constraints from Galaxy Clustering and Weak Lensing}

\input{DES-2017-0226_author_list.tex}
%%%\author{Dark Energy Survey Collaboration}

\date{\today}

%\pubyear{2017}

\label{firstpage}
\begin{abstract}
We present cosmological results from a combined analysis of galaxy clustering
and weak gravitational lensing, using 1321 deg$^2$ of $griz$ imaging data from
the first year of the Dark Energy Survey (DES Y1). We combine three two-point
functions: (i) the cosmic shear correlation function of 26 million source
galaxies in four redshift bins, (ii) the galaxy angular autocorrelation
function of 650,000 luminous red galaxies in five redshift bins, and (iii) the
galaxy-shear cross-correlation of luminous red galaxy positions and source
galaxy shears. To demonstrate the robustness of these results, we use
independent pairs of galaxy shape, photometric redshift estimation and
validation, and likelihood analysis pipelines. To prevent confirmation bias,
the bulk of the analysis was carried out while ``blind'' to the true results;
we describe an extensive suite of systematics checks performed and passed
during this blinded phase. The data are modeled in flat \lcdm\ and
\wcdm\ cosmologies, marginalizing over 20 nuisance parameters, varying 6 (for
\lcdm) or 7 (for \wcdm) cosmological parameters including the neutrino mass
density and including the 457 $\times$ 457 element analytic covariance matrix.
We find consistent cosmological results from these three two-point functions,
and from their combination obtain $S_8 \equiv \sigma_8 (\Omega_m/0.3)^{0.5} =
0.773^{+0.026}_{-0.020}$ and $\Omega_m = 0.267^{+0.030}_{-0.017}$ for \lcdm;
for \wcdm, we find $S_8 = 0.782^{+0.036}_{-0.024}$, $\Omega_m =
0.284^{+0.033}_{-0.030}$, and $w=-0.82^{+0.21}_{-0.20}$ at 68\% CL.  The
precision of these DES Y1 constraints rivals that from the Planck cosmic microwave
background measurements, allowing a comparison of structure in the very early
and late Universe on equal terms. Although the DES Y1 best-fit values for
$S_8$ and $\Omega_m$ are lower than the central values from Planck for both
\lcdm\ and \wcdm, the Bayes factor indicates that the DES Y1 and Planck data
sets are consistent with each other in the context of \lcdm.  Combining DES Y1
with Planck, Baryonic Acoustic Oscillation measurements from SDSS, 6dF, and
BOSS, and type Ia supernovae from the Joint Lightcurve Analysis (JLA) dataset,
we derive very tight constraints on cosmological parameters: $S_8 =
0.802\pm 0.012$ and $\Omega_m = 0.298\pm 0.007$ in \lcdm,
and $w = -1.00_{-0.04}^{+0.05}$ in \wcdm. Upcoming DES analyses will provide
more stringent tests of the \lcdm\ model and extensions such as a time-varying
equation of state of dark energy or modified gravity.
\vspace{25pt}

\end{abstract}

%\begin{keywords}
%gravitational lensing: weak -- cosmology: observations --
%surveys -- catalogues --
%methods: data analysis -- techniques: image processing
%\end{keywords}
\preprint{DES-2017-0226}
\preprint{FERMILAB-PUB-17-294-PPD}
\maketitle

\section{Introduction}
\label{sec:intro}

The discovery of cosmic acceleration \cite{riess98,perlmutter99} established
the Cosmological Constant ($\Lambda$)~\cite{Einstein:1917ce} + Cold Dark
 Matter (\lcdm) model as the standard cosmological paradigm that explains a
 wide variety of phenomena, from the origin and evolution of large-scale
 structure to the current epoch of accelerated expansion
 \cite{Lahav:2014vza,Weinberg:2012es}.  The successes of \lcdm, however, must
 be balanced by its apparent implausibility: three new entities beyond the
 Standard Model of particle physics --- one that drove an early epoch of
 inflation; another that serves as dark matter; and a third that is driving
 the current epoch of acceleration --- are required, none of them easily
 connected to the rest of physics \cite{Frieman:2008sn}. Ongoing and planned
 cosmic surveys are designed to test \lcdm\ and more generally to shed light
 on the mechanism driving the current epoch of acceleration, be it the vacuum
 energy associated with the cosmological constant, another form of dark
 energy, a modification of General Relativity, or something more drastic.

The Dark Energy Survey (DES\footnote{\url{http://www.darkenergysurvey.org/}},
\cite{Abbott:2005bi}) is an on-going, five-year survey that, when completed,
will map 300 million galaxies and tens of thousands of galaxy clusters in five
filters ($grizY$) over 5000 deg$^2$, in addition to discovering several
thousand type Ia supernovae in a 27 deg$^2$ time-domain survey.  DES will use
several cosmological probes to test \lcdm; galaxy clustering and weak
gravitational lensing are two of the most powerful. Jointly, these
complementary probes sample the underlying matter density field through the
galaxy population and the distortion of light due to gravitational lensing.
In this paper, we use data on this combination from the first year (Y1) of DES
to constrain \lcdm\ and its simplest extension---\wcdm, having a free
parameter for the dark energy equation of state.
 
The spatial distribution of galaxies in the Universe, and its temporal
evolution, carry important information about the physics of the early
Universe, as well as details of structure evolution in the late Universe,
thereby testing some of the most precise predictions of \lcdm.  Indeed,
measurements of the galaxy two-point correlation function, the lowest-order
statistic describing the galaxy spatial distribution, provided early evidence
for the \lcdm\ model
~\cite{Blumenthal:1984bp,Maddox:1990yw,Baugh:1995kz,Maddox:1996vz,Eisenstein:1999jg,1992MNRAS.254..295C,Szapudi:1997jp,Huterer:2000uj,Saunders:2000af,Hamilton:2000du,Cole:2005sx,Tegmark:2006az}. The
data--model comparison in this case depends upon uncertainty in the galaxy
{\it bias} ~\cite{Kaiser:1984sw}, the relation between the galaxy spatial
distribution and the theoretically predicted matter distribution.

In addition to galaxy clustering, weak gravitational lensing has become one of
the principal probes of cosmology.  While the interpretation of galaxy
clustering is complicated by galaxy bias, weak lensing provides direct
measurement of the mass distribution via \emph{cosmic shear}, the correlation
of the apparent shapes of pairs of galaxies induced by foreground large-scale
structure. Further information on the galaxy bias is provided by
\emph{galaxy--galaxy lensing}, the cross-correlation of lens galaxy positions
and source galaxy shapes.

The shape distortions produced by gravitational lensing, while cosmologically
informative, are extremely difficult to measure, since the induced source
galaxy ellipticities are at the percent level, and a number of systematic
effects can obscure the signal. Indeed, the first detections of weak lensing
were made by cross-correlating observed shapes of source galaxies with massive
foreground lenses~ \cite{Tyson:1990yt,Brainerd:1995da}. A watershed moment
came in the year 2000 when four research groups nearly simultaneously
announced the first detections of cosmic
shear~\cite{Bacon:2000sy,Kaiser:2000if,vanWaerbeke:2000rm,Wittman:2000tc}. While
these and subsequent weak lensing measurements are also consistent with \lcdm,
only recently have they begun to provide competitive constraints on
cosmological parameters
\cite{Jarvis:2005ck,Massey:2007gh,Schrabback:2009ba,Lin:2011bc,Heymans:2013fya,Huff:2011aa,kilbinger2013,Jee:2015jta,Hildebrandt:2016iqg,joudaki2017}.
Galaxy--galaxy lensing measurements have also matured to the point where their
combination with galaxy clustering breaks degeneracies between the
cosmological parameters and bias, thereby helping to constrain dark energy
\cite{Brainerd:1995da,Fischer:1999zx,Sheldon:2003xj,Leauthaud:2011rj,Mandelbaum:2005nx,Johnston:2007uc,Cacciato:2008hm,Mandelbaum:2012ay,Choi:2012kf,Velander:2013jga,Clampitt:2016ljk,Leauthaud:2016jdb,Kwan:2016mcy}.
The combination of galaxy clustering, cosmic shear, and galaxy--galaxy lensing
measurements powerfully constrains structure formation in the late universe.
As for cosmological analyses of samples of galaxy clusters \cite[see][for a
  review]{Allen:2011zs}, redshift space distortions in the clustering of
galaxies \cite[][and references therein]{Alam:2016hwk} and other measurements
of late-time structure, a primary test is whether these are consistent, in the
framework of \lcdm, with measurements from cosmic microwave background (CMB)
experiments that are chiefly sensitive to early-universe physics
\cite{Hinshaw:2012aka,Ade:2013zuv,Ade:2015xua,Calabrese:2017ypx} as well
  as lensing of its photons by the large-scale structures
  \cite[e.g.][]{Das:2011ak,vanEngelen:2012va,Ade:2015zua}.

The main purpose of this paper is to combine the information from galaxy
clustering and weak lensing, using the galaxy and shear correlation functions
as well as the galaxy-shear cross-correlation. It has been recognized for more
than a decade that such a combination contains a tremendous amount of
complementary information, as it is remarkably resilient to the presence of
nuisance parameters that describe systematic errors and non-cosmological
information
\cite{Hu:2003pt,Bernstein:2008aq,Joachimi:2009ez,Nicola:2016eua}. It is
perhaps simplest to see that the combined analysis could separately solve for
galaxy bias and the cosmological parameters; however, it can also internally
solve for (or, self-calibrate \cite{Huterer:2005ez}) the systematics
associated with photometric redshifts
\cite{Zhang:2009nc,Park:2015sxj,Samuroff:2016yqp}, intrinsic alignment
\cite{Zhang:2008pw}, and a wide variety of other effects
\cite{Joachimi:2009ez}.  Such a combined analysis has recently been executed
by combining the KiDS 450~deg$^{2}$ weak lensing survey with two different
spectroscopic galaxy surveys \cite{vanUitert:2017ieu,Joudaki:2017}.  While
these multi-probe analyses still rely heavily on prior information about the
nuisance parameters, obtained through a wide variety of physical tests and
simulations, this approach does significantly mitigate potential biases due to
systematic errors and will likely become even more important as statistical
errors continue to drop. The multi-probe analyses also extract more precise
information about cosmology from the data than any single measurement could.

Previously, the DES collaboration analyzed data from the Science Verification (SV) period, which covered
 139 deg$^2$, carrying out several pathfinding analyses of galaxy clustering and
gravitational lensing, along with numerous others \cite{Melchior:2014yap,Vikram:2015leg,Chang:2015odg,Becker:2015ilr,Abbott:2015swa,Crocce:2015xpb,Gruen:2015jhr,MacCrann:2016gac,Prat:2016xor,Clerkin:2016kyr,Chang:2016npo,Melchior:2016eyd,Kacprzak:2016vir,Kwan:2016mcy,Clampitt:2016ljk,Sanchez:2016mky,Abbott:2016ktf}. 
The DES Y1 data set analyzed here covers about ten times more area, albeit shallower, and provides
650,000 {\it lens} galaxies and the shapes of 26 million {\it source}
galaxies, each of them divided into redshift bins. 
The lens sample comprises bright, red-sequence galaxies, which have 
secure photometric redshift (photo-$z$) estimates.
We measure
three two-point functions from these data: (i) $w(\theta)$, the angular correlation function
of the lens galaxies; (ii) $\gamma_t(\theta)$, the correlation of the tangential shear of sources with lens galaxy positions; and (iii)
$\xi_\pm(\theta)$, the correlation functions of different components of the
ellipticities of the source galaxies.  We use these measurements only on large angular scales, for 
which we have verified that a relatively simple model describes the data, although
even with this restriction we must introduce twenty parameters to capture
astrophysical and measurement-related systematic uncertainties.

\newcommand\sims{\cite{simspaper}}
\newcommand\method{\cite{methodpaper}}
\newcommand\desdm{\cite{desdm}}
\newcommand\galaxy{\cite{y1gold}}
\newcommand\shape{\cite{shearcat}}
\newcommand\shearcorr{\cite{shearcorr}}
\newcommand\ggl{\cite{gglpaper}}
\newcommand\clustering{\cite{wthetapaper}}
\newcommand\pipeline{\cite{pipeline}}
\newcommand\photozpaper{\cite{photoz}}
\newcommand\redMaGiCpz{\cite{redMaGiCpz}}
\newcommand\xcorrtechnique{\cite{xcorrtechnique}}
\newcommand\xcorr{\cite{xcorr}}

This paper is built upon, and uses tools and results from, eleven other papers:
\begin{itemize}
\item Ref.~\method, which describes the theory and parameter-fitting methodologies, including the binning and modeling of all the two point functions, the marginalization of astrophysical and measurement related uncertainties, and the ways in which we calculate the covariance matrix and obtain the ensuing parameter constraints;
\item Ref.~\sims,  which applies this methodology to image simulations generated to mimic many aspects of the Y1 data sets;
\item a description of the process by which the value-added galaxy catalog (Y1 Gold) is created from the data and the tests on it 
to ensure its robustness~\galaxy;
\item a shape catalog paper, which presents the two shape catalogs
  generated using two independent techniques and the many tests
  carried out to ensure that residual systematic errors in the inferred shear estimates are sufficiently small
  for Y1 analyses~\shape;
\item Ref.~\photozpaper, which describes how the redshift distributions of galaxies in these shape catalogs are estimated from their photometry, including a validation of these estimates by means of COSMOS multi-band photometry;
\item three papers \cite{xcorrtechnique,redMaGiCpz,xcorr} that describe the use of angular cross-correlation with samples of secure redshifts to independently validate the photometric redshift distributions of lens and source galaxies;
\item Ref.~\shearcorr, which measures and derives cosmological constraints from the cosmic shear signal in the DES Y1 data and also addresses the question of whether DES lensing data are consistent with lensing results from other surveys;
\item Ref.~\ggl, which describes galaxy--galaxy lensing results, including a wide variety of tests for systematic contamination and a cross-check on the redshift distributions of source galaxies using the scaling of the lensing signal with redshift;
\item Ref.~\clustering, which describes the galaxy clustering statistics, including a series of tests for systematic contamination. This paper also describes updates to the redMaGiC algorithm used to select our lens galaxies and to estimate their photometric redshifts.
\end{itemize}

Armed with the above results, this paper presents the most stringent cosmological constraints from a galaxy imaging survey to date and, combined with external data, the most stringent constraints overall. 

One of the guiding principles of the methods developed in these papers is \emph{redundancy}: we use two independent shape measurement methods that are independently calibrated, several photometric redshift estimation and validation techniques, and two independent codes for predicting our signals and performing a likelihood analysis. Comparison of these, as described in the above papers, has been an important part of the verification of each step of our analysis.

The plan of the paper is as follows.  \S\ref{sec:data} gives an overview of
the data used in the analysis, while \S\ref{sec:twopoint} presents the
two-point statistics that contain the relevant information about cosmological
parameters. \S\ref{sec:method} describes the methodology used to compare these
statistics to theory, thereby extracting cosmological results. We validated
our methodology while remaining blinded to the results of the analyses; this
process is described in \S\ref{sec:blind}, and some of the tests that
convinced us to unblind are recounted in
Appendix~\ref{sec:unblind}. \S\ref{sec:des} presents the cosmological results
from these three probes as measured by DES in the context of two models,
\lcdm\ and \wcdm, while \S\ref{sec:ext} compares DES results with those from
other experiments, offering one of the most powerful tests to date of
\lcdm. Then, we combine DES with external data sets with which it is
consistent to produce the tightest constraints yet on cosmological
parameters. Finally, we conclude in
\S\ref{sec:con}. Appendix~\ref{sec:robustness} presents further evidence of
the robustness of our results. And Appendix~\ref{sec:covupdate} describes
updates in the covariance matrix calculation carried out after the first
version of this paper had been posted.

\section{Data}\label{sec:data}

DES uses the 570-megapixel Dark Energy Camera (DECam \cite{DECam}), built by
the collaboration and deployed on the Cerro Tololo Inter-American Observatory
(CTIO) 4m Blanco telescope in Chile, to image the South Galactic Cap in the
$grizY$ filters.  In this paper, we analyze DECam images taken from August 31,
2013 to February 9, 2014 (``DES Year 1'' or Y1), covering 1786 square degrees
in griz after coaddition and before masking \galaxy.  The data were processed
through the DES Data Management (DESDM) system
\citep{Desai:2012,Sevilla:2011,Mohr:2008,desdm}, which detrends and calibrates
the raw DES images, combines individual exposures to create coadded images,
and detects and catalogs astrophysical objects.  Further vetting and
subselection of the DESDM data products was performed by \galaxy\ to produce a
high-quality object catalog (Y1 Gold) augmented by several ancillary data
products including a star/galaxy separator. With up to 4 exposures per filter
per field in Y1, and individual $griz$ exposures of 90 sec and $Y$ exposures
of 45 sec, the characteristic 10$\sigma$ limiting magnitude for galaxies is
$g=23.4$, $r=23.2$, $i=22.5$, $z=21.8$, and $Y=20.1$ \galaxy.  Additional
analyses produced catalogs of red galaxies, photometric-redshift estimates,
and galaxy shape estimates, as described below.

As noted in \S\ref{sec:intro}, we use 
two samples of galaxies in the current analysis: {\it lens} 
galaxies, for the angular clustering measurement, and {\it source}
galaxies, whose shapes we estimate and correlate with each other
(``cosmic shear'').  The tangential shear is measured for the source galaxies about the positions
of the lens galaxies (galaxy--galaxy lensing).

\newcommand{\redMaGiC}{redMaGiC}
\newcommand{\zred}{z_{\rm red}}
\subsection{Lens Galaxies}

We rely on \redMaGiC\ galaxies for all galaxy clustering
measurements~\clustering ~and as the lens population for the galaxy--galaxy
lensing analysis~\ggl.  They have the advantage of being easily identifiable,
relatively strongly clustered, and of having relatively small
photometric-redshift errors; they are selected using a simple
algorithm~\cite{redmagicSV}:
\begin{enumerate}
\item Fit \textit{every} galaxy in the survey to a red-sequence template and
  compute the corresponding best-fit redshift $\zred$.
\item Evaluate the goodness-of-fit $\chi^2$ of the red-sequence template and
  the galaxy luminosity, using the assigned photometric redshift.
\item Include the galaxy in the \redMaGiC\ catalog if and only if it is bright
  $(L\geq L_{\rm min})$ and the red-sequence template is a good fit $(\chi^2
  \leq \chi^2_{\rm max})$.
\end{enumerate}
In practice, we do not specify $\chi^2_{\rm max}$ but instead demand that the
resulting galaxy sample have a constant comoving density as a function of
redshift.  Consequently, \redMaGiC\ galaxy selection depends upon only two
parameters: the selected luminosity threshold, $L_{\rm min}$, and the comoving
density, $\bar n$, of the sample.  Of course, not all combinations of
parameters are possible: brighter galaxy samples must necessarily be less
dense.

Three separate \redMaGiC\ samples were generated from the Y1 data, referred to
as the high-density, high-luminosity, and higher-luminosity samples.  The
corresponding luminosity thresholds\footnote{Here and throughout, whenever a cosmology is required, we use \lcdm\ with the parameters given in Table 1 of \method.} and comoving densities for these samples
are, respectively, $L_{\rm min} = 0.5L_*$, $L_*$, and $1.5L_*$, and $\bar n = 10^{-3}$,
$4\times 10^{-4}$,  and $10^{-4}\ {\rm galaxies}/(h^{-1} {\rm
  Mpc})^{3}$, where $h\equiv H_0/(100\, {\rm km}\,{\rm sec}^{-1}\,{\rm Mpc}^{-1}$) parametrizes the Hubble constant.  Naturally, brighter galaxies are easier to map at higher redshifts than are the dimmer galaxies.  These galaxies are placed  in five nominally disjoint redshift bins. The lowest three bins $z=[(0.15-0.3), (0.3-0.45), (0.45-0.6)]$ are high-density, while the galaxies in the two highest redshift bins ($(0.6-0.75)$ and $(0.75-0.9)$) are high-luminosity and higher-luminosity, respectively. The estimated redshift distributions of these five binned lens galaxy samples are shown in the upper panel of Figure~\rf{plot_nofzs}.

The clustering properties of these galaxies are an essential part of this
combined analysis, so great care is taken in \clustering\ to ensure that the
galaxy maps are not contaminated by systematic effects. This requires the
shallowest or otherwise irregular or patchy regions of the total 1786 deg$^2$
Y1 area to be masked, leaving a contiguous 1321 deg$^2$ as the area for the
analysis, the region called ``SPT'' in \cite{y1gold}.  The mask derived for
the lens sample is also applied to the source sample.

\subsection{Source Galaxies}

\subsubsection{Shapes}

Gravitational lensing shear is estimated from the statistical alignment of
shapes of source galaxies, which are selected from the Y1 Gold catalog
\galaxy. In DES Y1, we measure galaxy shapes and calibrate those measurements
by two independent and different algorithms, \metacal\ and \imshape, as
described in \shape.

\metacal\ \cite{Huff:2017qxu,Sheldon:2017szh} measures shapes by simultaneously fitting
  a 2D Gaussian model for each galaxy to the pixel data for all available $r$-, $i$-, and $z$-band exposures, convolving 
  with the point-spread functions (PSF) appropriate to each exposure.
This procedure is repeated on versions of these images that are artificially
sheared, i.e.~de-convolved, distorted by a shear operator, and re-convolved by
a symmetrized version of the PSF. By means of these, the \emph{response} of
the shape measurement to gravitational shear is measured from the images
themselves, an approach encoded in \metacal.

\metacal\ also includes an algorithm for calibration of shear-dependent
selection effects of galaxies, which could bias shear statistics at the few
percent level otherwise, by measuring on both unsheared and sheared images all
those galaxy properties that are used to select, bin and weight galaxies in
the catalog. Details of the practical application of these corrections to our
lensing estimators are given in
\cite{Sheldon:2017szh,shearcat,shearcorr,gglpaper}.

\imshape\ estimates a galaxy shape by determining the maximum likelihood set
of parameters from fitting either a bulge or a disc model to each object's
$r$-band observations \cite{Zuntz:2013zi}. The maximum likelihood fit, like
the Gaussian fit with \metacal, provides only a biased estimator of shear. For
\imshape, this bias is calibrated using a large suite of image simulations
that resemble the DES Y1 data set closely \cite{shearcat,des_sim_2017}.

Potential biases in the inferred shears are quantified by multiplicative
shear-calibration parameters $m^i$ in each source redshift bin $i$, such that
the measured shear $\gamma^{\rm meas} = (1+m^i) \gamma^{\rm true}$.  The $m^i$
are free parameters in the cosmological inferences, using prior constraints on
each as determined from the extensive systematic-error analyses in \shape.
These shear-calibration priors are listed in Table~\ref{tab:params}. The
overall \metacal\ calibration is accurate at the level of $1.3$ percent.  This
uncertainty is dominated by the impact of neighboring galaxies on shape
estimates.  For tomographic measurements, the widths of the overall $m^i$
prior is increased to yield a per-bin uncertainty in $m^i$, to account
conservatively for possible correlations of $m^i$ between bins \cite[see
  appendices of][]{shearcat,photoz}.  This yields the 2.3 percent prior per
redshift bin shown in Table~\ref{tab:params}.  The \imshape\ prior is
determined with 2.5 percent uncertainty for the overall sample (increased to a
3.5 percent prior per redshift bin), introduced mostly by imperfections in the
image simulations.

In both catalogs, we have applied conservative cuts, for instance on
signal-to-noise ratio and size, that reduce the number of galaxies with shape
estimates relative to the Y1 Gold input catalog significantly. For \metacal,
we obtain 35 million galaxy shape estimates down to an $r$-band magnitude of
$\approx 23$. Of these, 26 million are inside the restricted area and redshift
bins of this analysis.  Since its calibration is more secure, and its number
density is higher than that of \imshape\, (see \cite{shearcat} for details on
the catalog cuts and methodology details that lead to this difference in
number density), we use the \metacal\ catalog for our fiducial analysis.
\begin{table}
\caption{Parameters and priors\footnote{The lens \photoz\ priors changed
    slightly after unblinding due to changes in the cross-correlation
    analysis, as described in \redMaGiCpz; we checked that these changes did
    not impact our results.} used to describe the measured two-point
  functions. {\it Flat} denotes a flat prior in the range given while {\it
    Gauss}($\mu,\sigma$) is a Gaussian prior with mean $\mu$ and width
  $\sigma$. Priors for the tomographic nuisance parameters $m^{i}$ and $\Delta
  z^i$ have been widened to account for the correlation of calibration errors
  between bins \citep[][their appendix A]{photoz}. The $\Delta z^i$ priors
  listed are for \metacal\ galaxies and BPZ photo-$z$ estimates (see
  \cite{photoz} for other combinations). The parameter $w$ is fixed to $-1$ in
  the \lcdm\ runs.}
\begin{center}
\begin{tabular}{| c  c |}
\hline
\hline
Parameter & Prior \\  
\hline 
\multicolumn{2}{|c|}{{\bf Cosmology}} \\
$\Omega_m$  &  flat (0.1, 0.9)  \\ 
$A_s$ &  flat ($5\times 10^{-10},5\times 10^{-9}$)  \\ 
$n_s$ &  flat (0.87, 1.07)  \\
%$\wa$ &  0.0 &  flat (-2.5, 2.5)   \\
$\Omega_b$ &  flat (0.03, 0.07)  \\
$h$  &  flat (0.55, 0.91)   \\
$\Omega_\nu h^2$  & flat($5\times 10^{-4}$,$10^{-2}$) \\
$w$ &   flat ($-2$,$-0.33$)   \\
\hline
\multicolumn{2}{|c|}{{\bf Lens Galaxy Bias}} \\
$b_{i} (i=1,5)$   & flat (0.8, 3.0) \\
\hline
\multicolumn{2}{|c|}{{\bf Intrinsic Alignment}} \\
\multicolumn{2}{|c|}{{$A_{\rm IA}(z) = A_{\rm IA} [(1+z)/1.62]^\salpha$}} \\
$A_{\rm IA}$   & flat ($-5,5$) \\
$\salpha$   & flat ($-5,5$) \\
\hline
\multicolumn{2}{|c|}{{\bf Lens \photoz\ shift (red sequence)}} \\
$\Delta z^1_{\rm l}$  & Gauss ($0.008, 0.007$) \\
$\Delta z^2_{\rm l}$  & Gauss ($-0.005, 0.007$) \\
$\Delta z^3_{\rm l}$  & Gauss ($0.006, 0.006$) \\
$\Delta z^4_{\rm l}$  & Gauss ($0.000, 0.010$) \\
$\Delta z^5_{\rm l}$  & Gauss ($0.000, 0.010$) \\
\hline
\multicolumn{2}{|c|}{{\bf Source \photoz\ shift}} \\
$\Delta z^1_{\rm s}$  & Gauss ($-0.001, 0.016$) \\
$\Delta z^2_{\rm s}$  & Gauss ($-0.019, 0.013$) \\
$\Delta z^3_{\rm s}$  & Gauss ($+0.009, 0.011$) \\
$\Delta z^4_{\rm s}$  & Gauss ($-0.018, 0.022$) \\
\hline
\multicolumn{2}{|c|}{{\bf Shear calibration}} \\
$m^{i}_{\metacal} (i=1,4)$ & Gauss ($0.012, 0.023$)\\
$m^{i}_{\im} (i=1,4)$ & Gauss ($0.0, 0.035$)\\
\hline
\end{tabular}
\end{center}
\label{tab:params}
\end{table}

\subsubsection{Photometric redshifts}

Redshift probability distributions are also required for source galaxies in
cosmological inferences. For each source galaxy, the probability density that
it is at redshift $z$, $p_{\rm BPZ}(z)$, is obtained using a modified version
of the BPZ algorithm~\cite{2006AJ....132..926C}, as detailed in \photozpaper.
Source galaxies are placed in one of four redshift bins, $z=[(0.2-0.43),
  (0.43-0.63), (0.63-0.9), (0.9-1.3)]$, based upon the mean of their $p_{\rm
  BPZ}(z)$ distributions.  As described in~\photozpaper, \shearcorr~and~\ggl,
in the case of \metacal\ these bin assignments are based upon photo-$z$
estimates derived using photometric measurements made by the
\metacal\ pipeline in order to allow for correction of selection effects.

We denote by $n^{i}_{\rm PZ}(z)$ an initial estimate of the redshift
distribution of the $N^i$ galaxies in bin $i$ produced by randomly drawing a
redshift $z$ from the probability distribution $p_{\rm BPZ}(z)$ of each galaxy
assigned to the bin, and then bin all these $N^i$ redshifts into a
histogram. For this step, we use a BPZ estimate based on the optimal flux
measurements from the multi-epoch multi-object fitting procedure (MOF)
described in \cite{y1gold}.

For both the source and the lens galaxies, uncertainties in the redshift distribution are quantified by assuming that the 
true redshift distribution $n^i(z)$ in bin $i$ is a shifted version of the photometrically derived distribution:
\be
n^i(z) = n^{i}_{\rm PZ}(z - \Delta z^i),
\eql{nidef}
\ee
with the $\Delta z^i$ being free 
parameters in the cosmological analyses.  Prior constraints on these shift parameters are derived in two ways. 

First, we constrain $\Delta z^i$ from a matched sample of galaxies in the
COSMOS field, as detailed in~\photozpaper. Reliable redshift estimates for
nearly all DES-selectable galaxies in the COSMOS field are available from
30-band imaging \cite{Laigle:2016jxn}.  We select and weight a sample of
COSMOS galaxies representative of the DES sample with successful shape
measurements based on their color, magnitude, and pre-seeing size. The mean
redshift of this COSMOS sample is our estimate of the true mean redshift of
the DES source sample, with statistical and systematic uncertainties detailed
in~\photozpaper. The sample variance in the best-fit $\Delta z^i$ from the
small COSMOS field is reduced, but not eliminated, by reweighting the COSMOS
galaxies to match the multiband flux distribution of the DES source sample.

Second, the $\Delta z^i$ of both lens and source samples are further
constrained by the angular cross-correlation of each with a distinct sample of
galaxies with well-determined redshifts.  The $\Delta z^i_{\rm l}$ for the
three lowest-redshift lens galaxy samples are constrained by cross-correlation
of redMaGiC with spectroscopic redshifts~\redMaGiCpz\ obtained in the overlap
of DES Y1 with Stripe 82 of the Sloan Digital Sky Survey. The $\Delta z^i_{\rm
  s}$ for the three lowest-redshift source galaxy bins are constrained by
cross-correlating the sources with the redMaGiC sample, since the redMaGiC
photometric redshifts are much more accurate and precise than those of the
sources~\xcorrtechnique \xcorr.  The $z<0.85$ limit of the redMaGiC sample
precludes use of cross-correlation to constrain $\Delta z^4_{\rm s}$, so its
prior is determined solely by the reweighted COSMOS galaxies.

For the first three source bins, both methods yield an estimate of $\Delta
z^i_{\rm s}$, and the two estimates are compatible, so we combine them to
obtain a joint constraint. The priors derived for both lens and source
redshifts are listed in Table~\ref{tab:params}. The resulting estimated
redshift distributions are shown in Figure~\rf{plot_nofzs}.

Ref.~\cite{photoz} and Figure~\rf{dcosmos} in Appendix \ref{sec:robustness}
demonstrate that, at the accuracy attainable in DES Y1, the precise shapes of
the $n^i(z)$ functions have negligible impact on the inferred cosmology as
long as the mean redshifts of every bin, parametrized by the $\Delta z^i$, are
allowed to vary.  As a consequence, the cosmological inferences are
insensitive to the choice of photometric redshift algorithm used to establish
the initial $n^{i}_{\rm PZ}(z)$ of the bins.

\Sfig{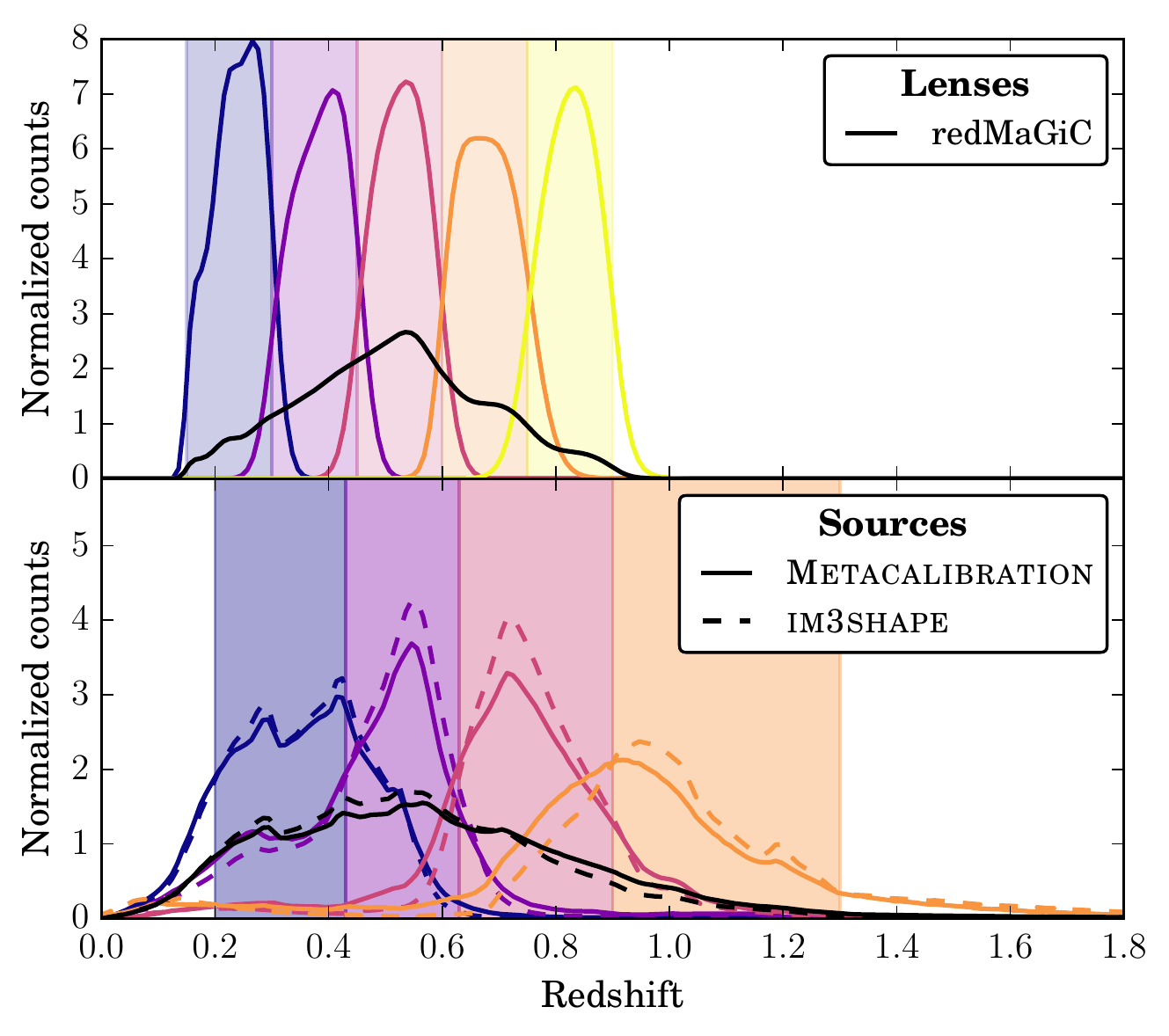}{Estimated redshift distributions of the lens and source
  galaxies used in the Y1 analysis. The shaded vertical regions define the
  bins: galaxies are placed in the bin spanning their mean
  \photoz\ estimate. We show both the redshift distributions of galaxies in
  each bin (colored lines) and their overall redshift distributions (black
  lines). Note that source galaxies are chosen via two different pipelines
  \im\ and \metacal, so their redshift distributions and total numbers differ
  (solid vs.~dashed lines).}

\section{Two-point Measurements}
\label{sec:twopoint}

We measure three sets of two-point statistics: the auto-correlation of the
positions of the \redMaGiC\ lens galaxies, the cross-correlation of the lens
positions with the shear of the source galaxies, and the two-point correlation
of the source galaxy shear field.  Each of the three classes of statistics is
measured using {\tt treecorr}~\cite{Jarvis:2003wq} in all pairs of redshift
bins of the galaxy samples and in 20 log-spaced bins of angular separation
$2.5^\prime<\theta<250^\prime,$ although we exclude some of the scales and
cross-correlations from our fiducial data vector (see
\autoref{sec:method}). Figures \rf{wgt} and \rf{xipm} show these measurements
and our best-fit \lcdm\ model.

\subsection{Galaxy Clustering: $w(\theta)$}

The inhomogeneous distribution of matter in the Universe is traced by
galaxies. The overabundance of pairs at angular separation $\theta$ above that
expected in a random distribution, $w(\theta)$, is one of the simplest
measurements of galaxy clustering.  It quantifies the strength and scale
dependence of the clustering of galaxies, which in turn reflects the
clustering of matter.

The upper panel of Figure~\rf{wgt} shows the angular correlation function of
the \redMaGiC\ galaxies in the five lens redshift bins described above. As
described in \clustering, these correlation functions were computed after
quantifying and correcting for spurious clustering induced by each of multiple
observational variables.  Figure~\rf{wgt} shows the data with the error bars
set equal to the square root of the diagonal elements of the covariance
matrix, but we note that data points in nearby angular bins are highly
correlated. Indeed, as can be seen in Figure~5 of \method, in the lowest
redshift bins the correlation coefficient between almost all angular bins is
close to unity; at higher redshift, the measurements are highly correlated
only over the adjacent few angular bins.  The solid curve in Figure~\rf{wgt}
shows the best-fit prediction from \lcdm\ after fitting to all three two-point
functions.  In principle, we could also use the angular cross-correlations
between galaxies in different redshift bins in the analysis, but the amount of
information in these cross-bin two-point functions is quite small and would
require substantially enlarging the covariance matrix, so we use only the
auto-correlations.

\subsection{Galaxy--galaxy lensing: $\gamma_t(\theta)$}

The shapes of background source galaxies are distorted by the mass associated
with foreground lenses. The characteristic distortion is a {\it tangential
  shear}, with the source galaxy ellipticities oriented perpendicular to the
line connecting the foreground and background galaxies. This shear,
$\gamma_t(\theta)$, is sensitive to the mass associated with the foreground
galaxies. On scales much larger than the sizes of parent halos of the
galaxies, it is proportional to the lens galaxy bias parameters $b^i$ in each
lens bin which quantifies the relative clumping of matter and galaxies. The
lower panels of Figure~\rf{wgt} show the measurements of galaxy--galaxy
lensing in all pairs of lens-source tomographic bins, including the model
prediction for our best-fit parameters.  The plots include bin pairs for which
the lenses are nominally behind the sources (those towards the upper right),
so might be expected to have zero signal. Although the signals for these bins
are expected to be small, they can still be useful in constraining the
intrinsic alignment parameters in our model (see, e.g.,
\cite{Troxel:2014dba}).

\Svwide{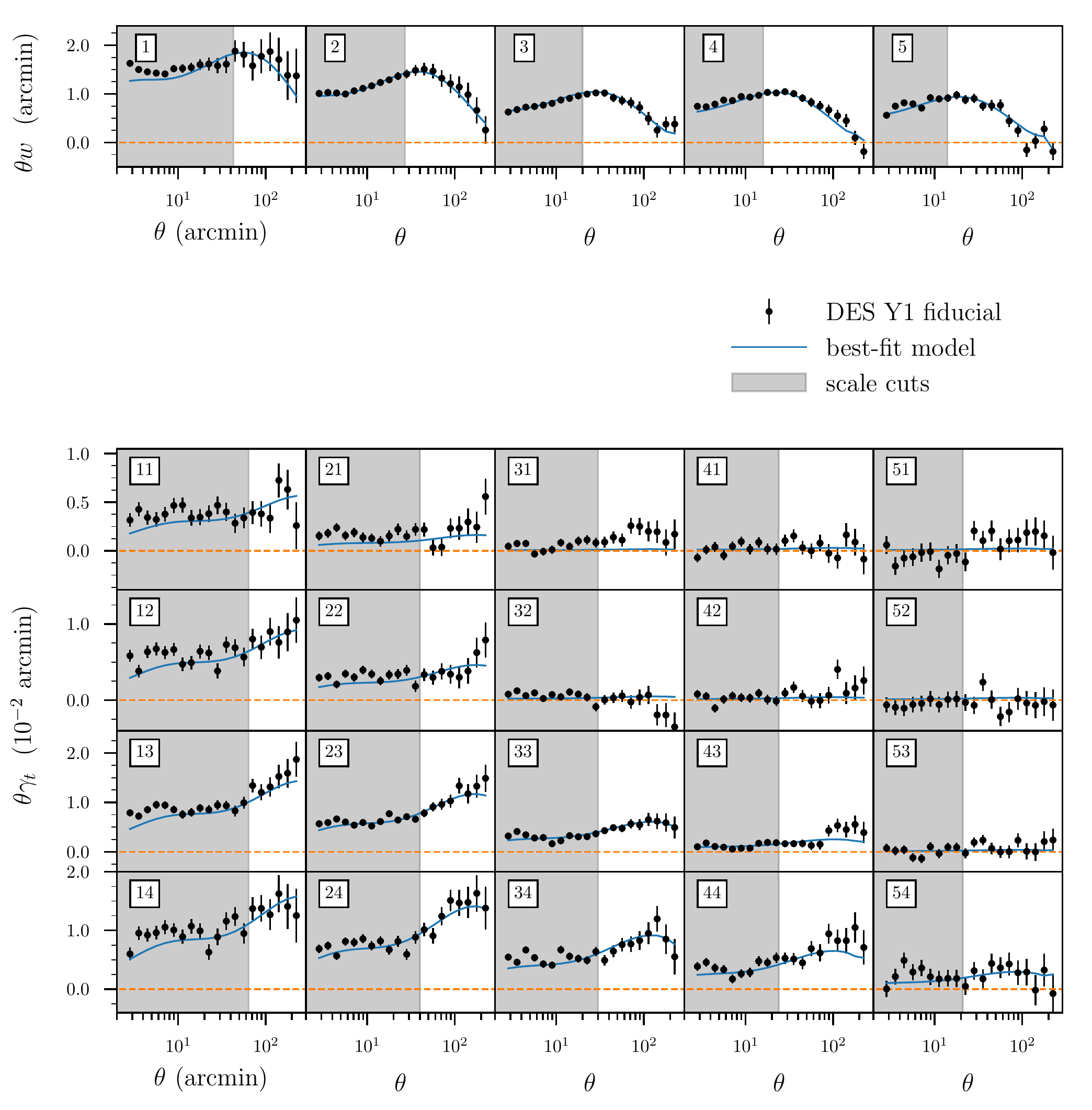}{Top panels: scaled angular correlation function, $\theta
  w(\theta)$, of \redMaGiC\ galaxies in the five redshift bins in the top
  panel of Figure \rf{plot_nofzs}, from lowest (left) to highest redshift
  (right) \clustering. The solid lines are predictions from the \lcdm\ model
  that provides the best fit to the combined three two-point functions
  presented in this paper. Bottom panels: scaled galaxy--galaxy lensing
  signal, $\theta \gamma_t$ (galaxy-shear correlation), measured in DES Y1 in
  four source redshift bins induced by lens galaxies in five \redMaGiC\ bins
  \ggl. Columns represent different lens redshift bins while rows represent
  different source redshift bins, so e.g., bin labeled 12 is the signal from
  the galaxies in the second source bin lensed by those in the first lens
  bin. The solid curves are again our best-fit \lcdm\ prediction. In all
  panels, shaded areas display the angular scales that have been excluded from
  our cosmological analysis (see \S\ref{sec:method}).}

In \ggl, we carried out a number of null tests to ensure the robustness of
these measurements, none of which showed evidence for significant systematic
uncertainties besides the ones characterized by the nuisance parameters in
this analysis.  The model fits the data well. Even the fits that appear quite
bad are misleading because of the highly off-diagonal covariance matrix.  For
the nine data points in the 3--1 bin, for example, $\chi^2=14,$ while $\chi^2$
would be 30 if the off-diagonal elements were ignored.

\subsection{Cosmic shear: $\xi_\pm(\theta)$}

The two-point statistics that quantify correlations between the shapes of
galaxies are more complex, because they are the products of the components of
a spin-2 tensor. Therefore, a pair of two-point functions are used to capture the
relevant information: $\xi_+(\theta)$ and $\xi_-(\theta)$ are the sum and
difference of the products of the tangential and cross components of the
shear, measured with respect to the line connecting each
galaxy pair. For more details, see \shearcorr\ or earlier work in
Refs~\cite{1991ApJ...380....1M,Kaiser:1991qi,Kaiser:1996tp,Kamionkowski:1997mp,Hui:1999ak,Bartelmann:1999yn,Refregier:2003ct,Hoekstra:2008db}. Figure~\rf{xipm}
shows these functions for different pairs of tomographic bins.

\Svwide{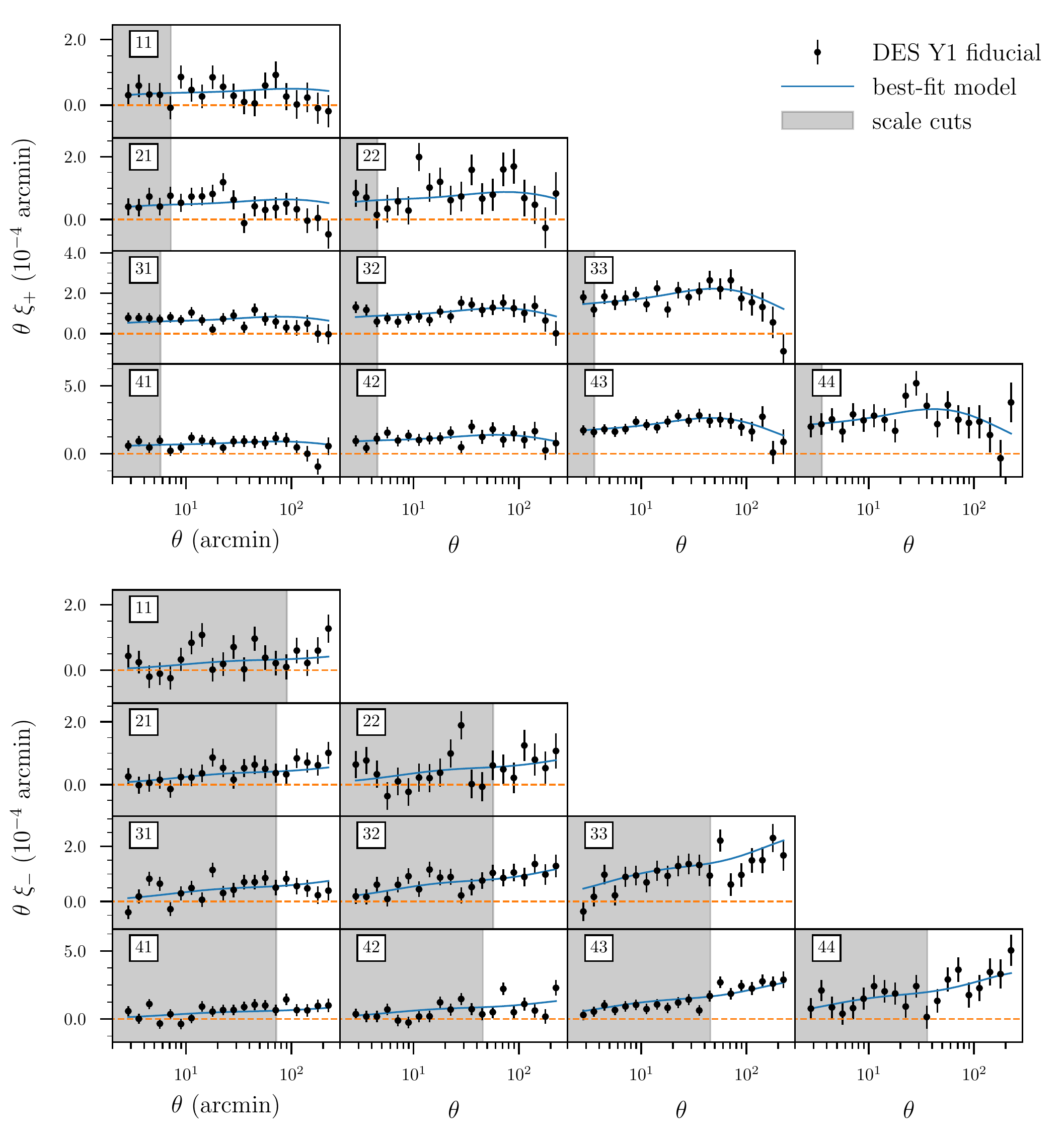}{The cosmic shear correlation functions $\xi_+$ (top
  panel) and $\xi_-$ (bottom panel) in DES Y1 in four source redshift
  bins, including cross correlations, measured from the \metacal\
  shear pipeline (see \shearcorr\ for the corresponding plot with
  \imshape); pairs of numbers in the upper left of each panel indicate the redshift bins. The solid lines show predictions from our best-fit \lcdm\ model from the analysis of all three two-point functions, and the shaded areas display the angular scales that are not used in our cosmological analysis (see \S\ref{sec:method}).}

As in Figure~\rf{wgt}, the best-fit model prediction here includes the impact
of intrinsic alignment; the best-fit shifts in the photometric redshift
distributions; and the best-fit values of shear calibration. The
one-dimensional posteriors on all of these parameters are shown in
Figure~\rf{table_1d_all} in Appendix \ref{sec:unblind}.

\section{Analysis}\label{sec:method}

\subsection{Model}

To extract cosmological information from these two-point functions, we construct a model that depends upon both cosmological parameters and astrophysical and observational nuisance parameters. The cosmological parameters govern the expansion history as well as the evolution and scale dependence of the matter clustering amplitude (as quantified, e.g., by the power spectrum). The nuisance parameters account for uncertainties in photometric redshifts, shear calibration, the bias between galaxies and mass, and the contribution of intrinsic alignment to the shear spectra. \S\ref{subsec:params} will enumerate these parameters, and our priors on them are listed in Table~\ref{tab:params}. Here, we describe how the two-point functions presented in \S\ref{sec:twopoint} are computed in the model.

\subsubsection{Galaxy Clustering: $w(\theta)$}
\label{sec:method:clustering}

Following \method, we express the projected (angular) density contrast of redMaGiC galaxies in redshift bin $i$ by $\delta_{\mathrm{g}}^i$, the convergence field of source tomography bin $j$ as $\kappa^j$, the redshift distribution of the redMaGiC/source galaxy sample in tomography bin $i$ as $n_{\mathrm{g}/\kappa}^i(z)$, and the angular number densities of galaxies in this redshift bin as 
\be
\bar{n}_{\mathrm{g}/\kappa}^i = \int dz\; n_{\mathrm{g}/\kappa}^i(z) \eql{nbar}\,.
\ee
The radial weight function for clustering in terms of the comoving radial distance $\chi$ is
\be
q_{\delta_{\mathrm{g}}}^i(k,\chi) = b^i\left(k,z(\chi)\right)\frac{n_{\mathrm{g}}^i(z(\chi)) }{\bar{n}_{\mathrm{g}}^i}\frac{dz}{d\chi} \eql{qdelta}\,,
\ee
with $b^i(k,z(\chi))$ the galaxy bias of the redMaGiC galaxies in tomographic bin $i$,
and the lensing efficiency 
\be
q_\kappa^{i}(\chi) = \frac{3 H_0^2 \Omega_m }{2 \mathrm{c}^2}\frac{\chi}{a(\chi)}\int_\chi^{\chi_{\mr h}} \mr d \chi' \frac{n_{\kappa}^{i} (z(\chi')) dz/d\chi'}{\bar{n}_{\kappa}^{i}} \frac{\chi'-\chi}{\chi'} \eql{qkappa} \,,
\ee
with $H_0$ the Hubble constant, $c$ the speed of light, and $a$ the scale factor.
Under the Limber approximation \cite{limber1954,kaiser1992,LoVerde:2008re,Kilbinger:2017lvu}, the angular correlation function for galaxy clustering can be written as
\bea
w^i(\theta) &=& \int \frac{dl\,l}{2\pi} J_0(l\theta)\int d\chi \, \frac{q_{\delta_{\mathrm{g}}}^i\!\!\left(\!\frac{l+1/2}{\chi},\chi\right)q_{\delta_{\mathrm{g}}}^j \!\left(\frac{l+1/2}{\chi},\chi\right)}{\chi^2}\vs
&&\times P_{\mathrm{NL}}\!\!\left(\frac{l+1/2}{\chi},z(\chi)\!\!\right) \eql{wtheta}
\eea
with $P_{\mathrm{NL}}(k,z)$ the non-linear matter power spectrum at wave vector $k$ and redshift $z$.

The expression in \ec{wtheta} and the ones in Eqs.~\Ec{gammat} and \Ec{xipm}
use the ``flat-sky'' approximation, which was tested against a curved sky
  implementation in \method\ for the case of galaxy clustering. Ref.~\method\ uses
  the more accurate expression that sums over Legendre polynomials, and we
  find that these two expressions show negligible differences over the scales
  of interest.

The model power spectrum here is the fully nonlinear power spectrum in
\lcdm\ or \wcdm, which we estimate on a grid of $(k,z)$ by first running {\tt
  CAMB}~\cite{Lewis:1999bs} or {\tt CLASS}~\cite{2011arXiv1104.2932L} to
obtain the linear spectrum and then {\tt
  HALOFIT}~\cite{smith2003,Takahashi:2012em,bird2012} for the nonlinear
spectrum.  The smallest angular separations for which the galaxy two-point
function measurements are used in the cosmological inference, indicated by the
boundaries of the shaded regions in the upper panels of Figure~\rf{wgt},
correspond to a comoving scale of 8$h^{-1}$ Mpc; this scale is chosen such
that modeling uncertainties in the non-linear regime cause negligible impact
on the cosmological parameters relative to their statistical errors, as shown
in \method\ and \shearcorr.

As described in \S{VI} of \method, we include the impact of neutrino
bias~\cite{bnu0,bnu1,bnu2} when computing the angular correlation function of
galaxies. For Y1 data, this effect is below statistical uncertainties, but it
is computationally simple to implement and will be relevant for upcoming
analyses.

\subsubsection{Galaxy--galaxy lensing: $\gamma_t(\theta)$}

We model the tangential shear similarly to how we modeled the angular correlation
  function. Consider the correlation of lens
  galaxy positions in bin $i$ with source galaxy shear in bin $j$; on large
  scales, it can be expressed as an integral over the power spectrum,
  \bea
  \gamma_{\mathrm t}^{ij}(\theta) &=& (1+m^j)\int
  \frac{dl\,l}{2\pi}\,J_2(l\theta)\, \int\!\! d\chi\!
  \frac{q_{\delta_{\mathrm{g}}}^i\!\!\left(\frac{l+1/2}{\chi},\chi\right)
    q_\kappa^j(\chi)}{\chi^2} \vs && \, \times
  P_{\mathrm{NL}}\!\left(\frac{l+1/2}{\chi},z(\chi)\right) \eql{gammat}
  \eea
where $m^j$ is the multiplicative shear bias, and $J_2$ is the 2nd-order Bessel
function.  The shift parameters characterizing the photo-z uncertainties
$\Delta z^j_{\rm s}$ and $\Delta z^i_{\rm l}$ enter the radial weight
functions in Eqs. \Ec{qdelta} and \Ec{qkappa} via Eqs. \Ec{nbar} and
\Ec{nidef}.    The shear signal also depends upon intrinsic alignments of the
source shapes with the tidal fields surrounding the lens galaxies; details of
our model for this effect (along with an examination of more complex models)
are given in \method\ and in \shearcorr. The smallest angular separations for
which the galaxy--galaxy lensing measurements are used in the cosmological
inference, indicated by the boundaries of the shaded regions in the lower
panels of Figure~\rf{wgt}, correspond to a comoving scale of 12$h^{-1}$ Mpc;
as above, this scale is chosen such that the model uncertainties in the
non-linear regime cause insignificant changes to the cosmological parameters
relative to the statistical uncertainties, as derived in ~\method\ and
verified in \sims.

\subsubsection{Cosmic shear $\xi_\pm(\theta)$}

The cosmic shear signal is independent of galaxy bias but shares the same general form as the other sets of two-point functions.
The theoretical predictions for these shear-shear two-point functions are
\bea
\xi_{+/-}^{ij}(\theta) &=& (1+m^i)(1+m^j)\int \frac{dl\,l}{2\pi}\,J_{0/4}(l\theta) \vs
&& \hspace{-0.5cm}\int d\chi  \frac{q_\kappa^i(\chi)q_\kappa^j(\chi)}{\chi^2}P_{\mathrm{NL}}\left(\frac{l+1/2}{\chi},z(\chi)\right)\eql{xipm}
\eea
where the efficiency functions are defined above, and $J_0$ and $J_4$ are the
Bessel functions for $\xi_+$ and $\xi_-$.
Intrinsic alignment affects the cosmic shear signal, especially the low-redshift bins, and are modeled as in \method. Baryons affect the matter power spectrum on small scales, and the cosmic shear signal is potentially sensitive to these uncertain baryonic effects; we restrict our analysis to the unshaded, large-scale regions shown in Figure~\rf{xipm} to reduce uncertainty in these effects below our measurement errors, following the analysis in \shearcorr.

\subsection{Parameterization and Priors}\label{subsec:params}

We use these measurements from the DES Y1 data to estimate cosmological
parameters in the context of two cosmological models, \lcdm\ and
\wcdm. \lcdm\ contains three energy densities in units of the critical
density: the matter, baryon, and massive neutrino energy densities, $\Omega_m,
\Omega_b,$ and $\Omega_\nu$. The energy density in massive neutrinos is a free
parameter but is often fixed in cosmological analyses to either zero or to a
value corresponding to the minimum allowed neutrino mass of 0.06 eV from
oscillation experiments \cite{pdg}. We think it is more appropriate to vary
this unknown parameter, and we do so throughout the paper (except in
\S\ref{sec:neutrinos}, where we show that this does not affect our qualitative
conclusions). We split the mass equally among the three eigenstates,
  hence assuming a degenerate mass hierarchy for the neutrinos.  Since most other survey
analyses have fixed $\Omega_\nu$, our results for the remaining parameters
will differ slightly from theirs, even when using their data.

\lcdm\ has three additional free parameters: the Hubble parameter, $H_0$, and
the amplitude and spectral index of the primordial scalar density
perturbations, $A_s$ and $n_s$. This model is based on inflation, which fairly
generically predicts a flat universe. Further when curvature is allowed to
vary in \lcdm, it is constrained by a number of experiments to be very close
to zero. Therefore, although we plan to study the impact of curvature in
future work, in this paper we assume the universe is spatially flat, with
$\Omega_\Lambda=1-\Omega_m$. It is common to replace $A_s$ with the RMS
amplitude of mass fluctuations on 8 $h^{-1}$ Mpc scale in linear theory,
$\sigma_8$, which can be derived from the aforementioned parameters. Instead
of $\sigma_8$, in this work we will focus primarily on the related parameter
\begin{equation}
S_8\equiv \sigma_8 \left( \frac{\Omega_m}{0.3}\right)^{0.5} \eql{s8}
\end{equation}
since $S_8$ is better constrained than $\sigma_8$ and is
largely uncorrelated with $\Omega_m$ in the DES parameter posterior.

We also consider the possibility that the dark energy is not a cosmological
constant. Within this \wcdm\ model, the dark energy equation of state
parameter, $w$ (not to be confused with the angular correlation function
$w(\theta)$), is taken as an additional free parameter instead of being fixed
at $w=-1$ as in \lcdm. \wcdm\ thus contains 7 cosmological parameters. In
future analyses of larger DES data sets, we anticipate constraining more
extended cosmological models, e.g., those in which $w$ is allowed to vary in
time.

In addition to the cosmological parameters, our model for the data contains 20
nuisance parameters, as indicated in the lower portions of
Table~\ref{tab:params}. These are the nine shift parameters, $\Delta z^i$, for
the source and lens redshift bins, the five \redMaGiC\ bias parameters, $b^i$,
the four multiplicative shear biases, $m^i$, and two parameters, $A_{\rm IA}$
and $\salpha$, that parametrize the intrinsic alignment model.

Table \ref{tab:params} presents the priors we impose on the cosmological and
nuisance parameters in the analysis. For the cosmological parameters, we
generally adopt flat priors that span the range of values
well beyond the uncertainties reported by recent experiments. As an example,
although there are currently potentially conflicting measurements of $h$, we
choose the lower end of the prior to be 10$\sigma$ below the lower central
value from the Planck cosmic microwave background
measurement~\cite{Ade:2015xua} and the upper end to be 10$\sigma$ above the
higher central value from local measurements~\cite{Riess:2016jrr}. In the case
of \wcdm, we impose a physical upper bound of $w<-0.33$, as that is required
to obtain cosmic acceleration. As another example, the lower bound of the
prior on the massive neutrino density, $\Omega_\nu h^2$, in
Table~\ref{tab:params} corresponds to the experimental lower limit on the sum
of neutrino masses from oscillation experiments.

For the astrophysical parameters $b^i$, $A_{\rm IA}$, and $\salpha$ that are
not well constrained by other analyses, we also adopt conservatively wide,
flat priors. For all of these relatively uninformative priors, the guiding
principle is that they should not impact our final results, and in particular
that the tails of the posterior parameter distributions should not lie close
to the edges of the priors\footnote{The sole exception is the
    intrinsic-alignment parameter $\salpha$ for which the posterior does hit
    the edge of the (conservatively selected, given feasible IA evolution) prior; see Figure~\rf{table_1d_all}
    in the Appendix~\ref{sec:unblind}.}. For the remaining nuisance parameters, $\Delta z^i$
and $m^i$, we adopt Gaussian priors that result from the comprehensive
analyses described in
Refs.~\cite{photoz,redMaGiCpz,shearcat,xcorrtechnique,xcorr}. The prior and
posterior distributions of these parameters are plotted in
Appendix~\ref{sec:unblind} in Figure~\rf{table_1d_all}.

In evaluating the likelihood function (\S\ref{subsec:like}), the parameters
with Gaussian priors are allowed to vary over a range roughly five times wider
than the prior; for example, the parameter that accounts for a possible shift
in the furthest lens redshift bin, $\Delta z_{\rm l}^5$, has a 1-$\sigma$
uncertainty of 0.01, so it is allowed to vary over $\vert \Delta z_{\rm
  l}^5\vert < 0.05$. These sampling ranges conservatively cover the parameter
values of interest while avoiding computational problems associated with
exploring parameter ranges that are overly broad. Furthermore, overly broad
parameter ranges would distort the computation of the Bayesian evidence, which
would be problematic as we will use Bayes factors to assess the consistency of
the different two-point function measurements, consistency with external data
sets, and the need to introduce additional parameters (such as $w$) into the
analysis. We have verified that our results below are insensitive to the
prior ranges chosen.

\subsection{Likelihood Analysis}\label{subsec:like}

For each data set, we sample the likelihood, assumed to be Gaussian, in the
many-dimensional parameter space:
\begin{equation}
\ln\mathcal{L}(\vec p) = -\frac{1}{2}\sum_{ij} \left[D_i-T_i(\vec p)\right] C^{-1}{}_{ij} \left[D_j-T_j(\vec p)\right] ,\eql{likeli}
\end{equation}
where $\vec p$ is the full set of parameters, $D_i$ are the measured two-point
function data presented in Figures~\rf{wgt} and \rf{xipm}, and $T_i(\vec p)$
are the theoretical predictions as given in \eeec{wtheta}{gammat}{xipm}. The
likelihood depends upon the covariance matrix $C$ that describes how the
measurement in each angular and redshift bin is correlated with every other
measurement. Since the DES data vector contains $457$ elements, the covariance
is a symmetric $457\times 457$ matrix. We generate the covariance matrices
using {\tt CosmoLike}~\cite{Krause:2016jvl}, which computes the relevant
four-point functions in the halo model, as described in \method. We also
describe there how the {\tt CosmoLike}-generated covariance matrix is tested
with simulations.

\ec{likeli} leaves out the $\ln(\det(C))$ in the prefactor\footnote{However,
  this factor is important for the Bayesian evidence calculations discussed
  below so is included in those calculations.} and more generally neglects the
cosmological dependence of the covariance matrix. Previous
work~\cite{Eifler:2008gx} has shown that this dependence is likely to have a
small impact on the central value; our rough estimates of the impact of
neglecting the determinant confirm this; and --- as we will show below --- our
results did not change when we replaced the covariance matrix with an updated
version based on the best-fit parameters. However, as we will see, the
uncertainty in the covariance matrix leads to some lingering uncertainty in
the error bars. To form the posterior, we multiply the likelihood by the
priors, $\mathcal{P}(\vec p)$, as given in Table~\ref{tab:params}.

Parallel pipelines, {\tt CosmoSIS}\footnote{{\tt
    https://bitbucket.org/joezuntz/cosmosis/}}~\cite{Zuntz:2014csq} and {\tt
  CosmoLike}, are used to compute the theoretical predictions and to generate
the Monte Carlo Markov Chain (MCMC) samples that map out the posterior space
leading to parameter constraints.  The two sets of software use the publicly
available samplers {\tt MultiNest}~\cite{Feroz:2008xx} and {\tt
  emcee}~\cite{ForemanMackey:2012ig}. The former provides a powerful way to
compute the Bayesian evidence described below so most of the results shown
here use {\tt CosmoSIS} running {\tt MultiNest}.

\subsection{Tests on Simulations}

The collaboration has produced a number of realistic mock catalogs for the DES
Y1 data set, based upon two different cosmological $N$-body simulations
(Buzzard~\cite{busha13}, MICE~\cite{fosalba15a}), which were analyzed as
described in \sims. We applied all the steps of the analysis on the
simulations, from measuring the relevant two-point functions to extracting
cosmological parameters. In the case of simulations, the true cosmology is
known, and \sims\ demonstrates that the analysis pipelines we use here do
indeed recover the correct cosmological parameters.

\section{Blinding and Validation}\label{sec:blind}

The small statistical uncertainties afforded by the Y1 data set present an
opportunity to obtain improved precision on cosmological parameters, but also
a challenge to avoid confirmation biases. To preclude such biases, we followed
the guiding principle that decisions on whether the data analysis has been
successful should not be based upon whether the inferred cosmological
parameters agreed with our previous expectations. We remained blind to the
cosmological parameters implied by the data until after the analysis procedure
and estimates of uncertainties on various measurement and astrophysical
nuisance parameters were frozen.

To implement this principle, we first transformed the ellipticities $e$ in the
shear catalogs according to $\arctanh\vert e\vert\rightarrow \lambda
\arctanh\vert e\vert,$ where $\lambda$ is a fixed blind random number between
0.9 and 1.1.  Second, we avoided plotting the measured values and theoretical
predictions in the same figure (including simulation outputs as
``theory''). Third, when running codes that derived cosmological parameter
constraints from observed statistics, we shifted the resulting parameter
values to obscure the best-fit values and/or omitted axis labels on any plots.

These measures were all kept in place until the following criteria were satisfied:
\begin{enumerate}

\item All non-cosmological systematics tests of the shear measurements were
  passed, as described in \shape, and the priors on the multiplicative biases
  were finalized.
\item Photo-$z$ catalogs were finalized and passed internal tests, as
  described in \cite{photoz,xcorrtechnique,redMaGiCpz,xcorr}.
\item Our analysis pipelines and covariance matrices, as described in
  \cite{simspaper,methodpaper}, passed all tests, including robustness to
  intrinsic alignment and bias model assumptions.
\item We checked that the \lcdm\ constraints (on, e.g., $\Omega_m,\sigma_8$)
  from the two different cosmic shear pipelines \im\ and \metacal\ agreed. The
  pipelines were \emph{not} tuned in any way to force agreement.
\item \lcdm\ constraints were stable when dropping the smallest angular bins
  for \metacal\ cosmic shear data. \label{shearonly}
\item Small-scale \metacal\ galaxy--galaxy lensing data were consistent
  between source bins (shear-ratio test, as described in \S 6 of \ggl). We
  note that while this test is performed in the nominal \lcdm\ model, it is
  close to insensitive to cosmological parameters, and therefore does not
  introduce confirmation bias.  \setcounter{syscounter}{\value{enumi}}
\end{enumerate}

Once the above tests were satisfied, we unblinded the shear catalogs but kept
cosmological parameter values blinded while carrying out the following checks,
details of which can be found in Appendix~\ref{sec:unblind}:
\begin{enumerate}
\setcounter{enumi}{\value{syscounter}}
\item Consistent results were obtained from the two theory/inference
  pipelines, {\tt CosmoSIS} and {\tt CosmoLike}.
\label{cs_vs_cl}
\label{posteriors_vs_priors}
\item Consistent results on all cosmological parameters were obtained with the
  two shear measurement pipelines, \metacal\ and \imshape.
\label{metacal_vs_imshape}
\item Consistent results on the cosmological parameters were obtained when we
  dropped the smallest-angular-scale components of the data vector, reducing
  our susceptibility to baryonic effects and departures from linear galaxy
  biasing.  This test uses the combination of the three two-point functions
  (as opposed to from shear only as in test~\ref{shearonly}).
\label{fiducial_vs_largescale}
\item An acceptable goodness-of-fit value ($\chi^2$) was found between the
  data and the model produced by the best-fitting parameters.  This assured us
  that the data were consistent with \emph{some} point in the model space that
  we are constraining, while not yet revealing which part of parameter space
  that is.
\label{chisq}
\item Parameters inferred from cosmic shear ($\xi_\pm$) were consistent with
  those inferred from the combination of galaxy--galaxy lensing ($\gamma_t$)
  and galaxy clustering ($w(\theta)$).
\label{cs_vs_2x2}
\end{enumerate}

Once these tests were satisfied, we unblinded the parameter inferences.  The
following minor changes to the analysis procedures or priors were made after
the unblinding: as planned before unblinding, we re-ran the MCMC chains with a
new covariance matrix calculated at the best-fit parameters of the original
analysis.  This did not noticeably change the constraints (see
Figure~\rf{dcov} in Appendix~\ref{sec:robustness}), as expected from our
earlier tests on simulated data \cite{methodpaper}.  We also agreed before
unblinding that we would implement two changes after unblinding: small changes
to the photo-$z$ priors referred to in the footnote to Table~\ref{tab:params},
and fixing a bug in \imshape\ object blacklisting that affected $\approx1\%$
of the footprint.

All of the above tests passed, most with reassuringly unremarkable results;
more details are given in Appendix~\ref{sec:unblind}.

For test \ref{chisq}, we calculated the $\chi^2$ ($=-2\log L$) value of the
457 data points used in the analysis using the full covariance matrix. In
\lcdm, the model used to fit the data has 26 free parameters, so the number of
degrees of freedom is $\nu=431$.  The model is calculated at the best-fit
parameter values of the posterior distribution (i.e. the point from the
posterior sample with lowest $\chi^2$). Given the uncertainty on the estimates
of the covariance matrix, the formal probabilities of a $\chi^2$ distribution
are not applicable. We agreed to unblind as long as $\chi^2$ was less than 605
($\chi^2/\nu<1.4$).  The best-fit value $\chi^2=497$ passes this test
 \footnote{In our original analysis (submitted to the arXiv in August
     2017), we originally found $\chi^2=572$, which passed the aforementioned
     criterion ($\chi^2<605$) with proceeding in the analysis. We have since
     identified a couple of missing ingredients in our computation of the
     covariance matrix, leading to the present, lower, value
     $\chi^2=497$. While the chi squared has significantly decreased, the
     cosmological constraints are nearly unchanged. Please see Appendix
     \ref{sec:covupdate} for more details.}, with $\chi^2/\nu=1.16.$
 Considering the fact that 13 of the free parameters are nuisance parameters
 with tight Gaussian priors, we will use $\nu=444,$ giving
 $\chi^2/\nu=1.12.$

The best-fit models for the three two-point functions are over-plotted on the
data in Figures~\rf{wgt} and \rf{xipm}, from which it is apparent that the
$\chi^2$ is not dominated by conspicuous outliers. Figure~\rf{bfhist} offers
confirmation of this, in the form of a histogram of the differences between
the best-fit theory and the data in units of the standard deviation of
individual data points.  The three probes show similar values of $\chi^2/\nu$:
for $\xi_\pm(\theta),$ $\chi^2 = 230$ for 227 data points; for
$\gamma_t(\theta),$ $\chi^2 = 185$ for 176 data points;
and for $w(\theta),$ $\chi^2 = 68$ for 54 data points. A finer
division into each of the 45 individual 2-point functions shows no significant
concentration of $\chi^2$ in particular bin pairs. We also find that removing
all data at scales $\theta>100^\prime$ yields $\chi^2= 278$ for 277 data
points ($\chi^2/\nu=1.05$), not a significant reduction, and also yields
no significant shift in best-fit parameters. Thus, we find that no particular
piece of our data vector dominates our $\chi^2$ result.
 
\Sfig{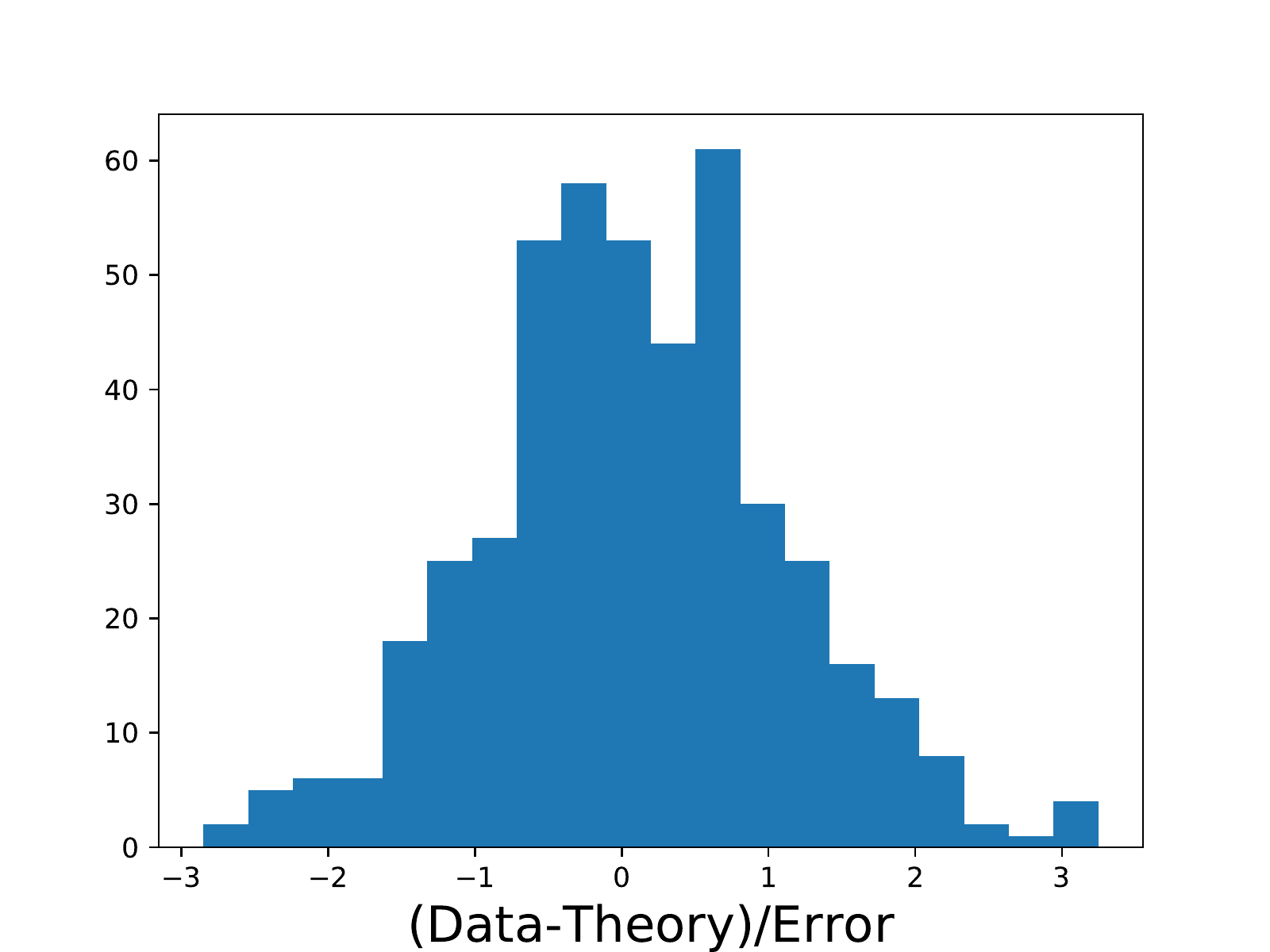}{Histogram of the differences between the best-fit \lcdm\ model
  predictions and the 457 data points shown in Figures \rf{wgt} and \rf{xipm},
  in units of the standard deviation of the individual data points. Although
  the covariance matrix is not diagonal, and thus the diagonal error bars do
  not tell the whole story, it is clear that there are no large outliers that
  drive the fits. }

Finally, for step number \ref{cs_vs_2x2} in the test list near the
  beginning of this Section, we examined several measures of consistency
between (i) cosmic shear and (ii) $\gamma_t(\theta)+w(\theta)$ in \lcdm.  As
an initial test, we computed the mean of the 1D posterior distribution of each
of the cosmological parameters and measured the shift between (i) and (ii).
We then divided this difference by the expected standard deviation of this
difference (taking into account the estimated correlation between the
$\xi_\pm$ and $\gamma_t+w$ inferences), $\sigma_{\rm
    diff}=[\sigma_{\xi_\pm}^2+\sigma_{\gamma_tw}^2-2\mathrm{Cov}(\xi_\pm,
    \gamma_t+w)]^{1/2}$.  For all parameters, these differences had absolute
value $<0.4$, indicating consistency well within measurement error.

As a second consistency check, we compared the posteriors for the nuisance
parameters from cosmic shear to those from clustering plus galaxy--galaxy
lensing, and they agreed well. We found no evidence that any of the nuisance
parameters pushes against the edge of its prior or that the nuisance
parameters for cosmic shear and $w+\gamma_t$ are pushed to significantly
different values. The only mild exceptions are modest shifts in the intrinsic
alignment parameters, $A_{\rm IA}$ and $\salpha$, as well as in the
second source redshift bin, $\Delta z_s^2$. The full set of posteriors on all
20 nuisance parameters for \metacal\ is shown in Figure~\rf{table_1d_all} in
Appendix~\ref{sec:unblind}.

As a final test of consistency between the two sets of two-point-function
measurements, we use the Bayes factor (also called the ``evidence ratio'').
The Bayes factor is used for discriminating between two hypotheses, and is the
ratio of the Bayesian evidences, $P(\vec D | H)$ (the probability of observing
dataset, $\vec D$, given hypothesis $H$) for each hypothesis. An example of
such a hypothesis is that dataset $\vec D$ can be described by a model $M$, in
which case the Bayesian evidence is
\begin{equation}
P(\vec D | H) = \int d^N \theta P(\vec D | \vec\theta, M) P(\vec \theta | M)
\end{equation}
where $P(\vec D | \vec\theta, M)$ is the likelihood of the data given the
  model $M$ parametrized by its $N$ parameters $\vec \theta$, and $P(\vec
  \theta | M)$ is the prior probability distribution of those model  parameters.

For two hypotheses $H_0$ and $H_1$, the Bayes factor is given by
\begin{equation}
R = \frac{P(\vec D | H_0)}{P(\vec D | H_1)} = \frac{P(H_0 |\vec D) P(H_1)}{P(H_1 | \vec D) P(H_0)}\eql{}
\end{equation}
where the second equality follows from Bayes' theorem and clarifies the
meaning of the Bayes factor: if we have equal {\it a priori belief} in $H_0$
and $H_1$ (i.e., $P(H_0)=P(H_1)$), the Bayes factor is the ratio of the
posterior probability of $H_0$ to the posterior probability of $H_1$. The
Bayes factor can be interpreted in terms of odds, i.e., it implies $H_0$ is
favored over $H_1$ with $R:1$ odds (or disfavored if $R<1$). We will adopt
the widely used Jeffreys scale~\cite{jeffreys61} for interpreting Bayes
factors: $3.2<R<10$ and $R>10$ are respectively considered \emph{substantial}
and \emph{strong} evidence for $H_0$ over $H_1$. Conversely, $H_1$ is strongly
favored over $H_0$ if $R<0.1$, and there is substantial evidence for $H_1$ if
$0.1<R<0.31$.
 %On the Kass and Raftery scale~\cite{kassraftery95},
 %$3<R<20$ and $R>20$ are considered \emph{positive} and \emph{strong} evidence for $H_0$ over $H_1$.
 
We follow \cite{marshall06} by applying this formalism as a test for consistency
between cosmological probes. In this case, the null hypothesis, $H_0$, is that the two datasets were 
measured from the same universe and therefore share the same model parameters. Two probes would be judged discrepant if they strongly favor the alternative hypothesis, $H_1$, 
that they are measured from two different universes with different model parameters. 
So the appropriate Bayes factor for judging consistency of two datasets, $D_1$ and $D_2$, is
\be
R = \frac{ P\left(\vec D_1, \vec D_2 \vert M\right) }{P\left(\vec D_1 \vert M\right) P\left(\vec D_2 \vert M\right)}\eql{evidence}
\ee
where $M$ is the model, e.g., \lcdm\ or \wcdm.
The numerator is the evidence for both datasets when model $M$ is fit to both datasets simultaneously. The denominator is the evidence for both datasets when model M is fit to both datasets individually, and therefore each dataset determines its own parameter posteriors. 

Before the data were unblinded, we decided that we would combine results from
these two sets of two-point functions if the Bayes factor defined in
\ec{evidence} did not suggest strong evidence for inconsistency. According to
the Jeffreys scale, our condition to combine is therefore that $R>0.1$ (since
$R<0.1$ would imply strong evidence for inconsistency).  We find a Bayes
factor of $R=583$, an indication that DES Y1 cosmic shear and galaxy
clustering plus galaxy--galaxy lensing are consistent with one another in the
context of \lcdm.

The DES Y1 data were thus validated as internally consistent and robust to our
assumptions before we gained any knowledge of the cosmological parameter
values that they imply.  Any comparisons to external data were, of course,
made after the data were unblinded.

\section{DES Y1 Results: Parameter Constraints}\label{sec:des}

\subsection{\lcdm}

We first consider the  \lcdm\ model with six cosmological parameters. The DES data are most sensitive to two cosmological parameters, $\Omega_m$ and $S_8$ as defined in \ec{s8}, so for the most part we focus on constraints on these parameters.

\Sfig{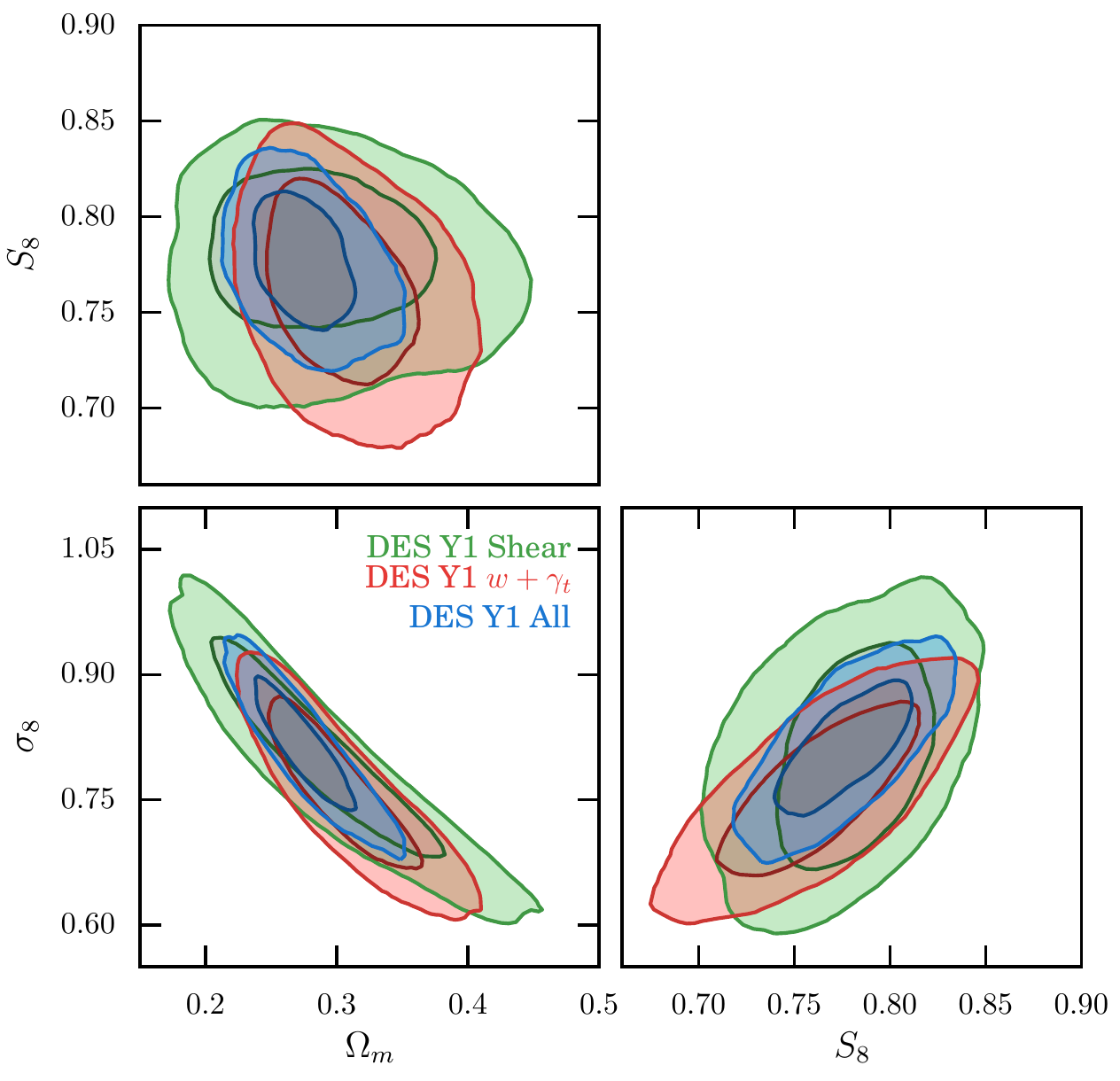}{\lcdm\ constraints from DES Y1 on $\Omega_m, \sigma_8$, and $S_8$
  from cosmic shear (green), \redMaGiC\ galaxy clustering plus galaxy--galaxy
  lensing (red), and their combination (blue).  Here, and in all such 2D plots
  below, the two sets of contours depict the 68\% and 95\% confidence levels.
}

Given the demonstrated consistency of cosmic shear with clustering plus
galaxy--galaxy lensing in the context of \lcdm\ as noted above, we proceed to
combine the constraints from all three probes.  Figure~\rf{ss8} shows the
constraints on $\Omega_m$ and $\sigma_8$ (bottom panel), and on $\Omega_m$ and
the less degenerate parameter $S_8$ (top panel). Constraints from cosmic
shear, galaxy clustering + galaxy--galaxy lensing, and their combination are
shown in these two-dimensional subspaces after marginalizing over the 24 other
parameters. The combined results lead to constraints
\bea
\Omega_m &=& 0.267^{+0.030}_{-0.017}\vs
S_8 &=& 0.773^{+0.026}_{-0.020}\vs
\sigma_8 &=& 0.817_{-0.056}^{+0.045}
.\eea

The value of $\Omega_m$ is consistent with the value inferred from
either cosmic shear or clustering plus galaxy--galaxy lensing separately.  We
present the resulting marginalized constraints on the cosmological parameters
in the top rows of Table~\ref{tab:post}.

\begin{table*}
\setlength{\extrarowheight}{7pt}
\caption{\label{tab:post} 68\%CL marginalized cosmological constraints in
  \lcdm\ and \wcdm\ using a variety of datasets. ``DES Y1 3x2'' refers to
  results from combining all 3 two-point functions in DES Y1. Cells with no
  entries correspond to posteriors not significantly narrower than the prior
  widths. The only exception is in \wcdm\ for Planck only, where the
  posteriors on $h$ are shown to indicate the large values inferred in the
  model without any data to break the $w-h$ degeneracy.}
\begin{tabular}{|c|c|c|c|c|c|c|c|c|}
\hline
Model & Data Sets&$\Omega_m$ &$S_8$ &$n_s$ &$\Omega_b$ &$h$ &\parbox{1.6cm}{  $\sum m_\nu$ (eV) \\ (95\% CL)  } &$w$ \\ 
\hline
\hline
$\Lambda$CDM\ & DES Y1 $\xi_\pm(\theta)$ & $0.260_{-0.037}^{+0.065}$ & $0.782_{-0.027}^{+0.027}$ & $\ldots$ & $\ldots$ & $\ldots$ & $\ldots$ & $\ldots$ \\ 
\hline
$\Lambda$CDM\ & DES Y1 $w(\theta)+\gamma_t$ & $0.288_{-0.026}^{+0.045}$ & $0.760_{-0.030}^{+0.033}$ & $\ldots$ & $\ldots$ & $\ldots$ & $\ldots$ & $\ldots$ \\ 
\hline
$\Lambda$CDM\ & DES Y1 3x2 & $0.267_{-0.017}^{+0.030}$ & $0.773_{-0.020}^{+0.026}$ & $\ldots$ & $\ldots$ & $\ldots$ & $\ldots$ & $\ldots$ \\ 
\hline
$\Lambda$CDM\ & Planck (No Lensing) & $0.334_{-0.026}^{+0.037}$ & $0.841_{-0.025}^{+0.027}$ & $0.958_{-0.005}^{+0.008}$ & $0.0503_{-0.0019}^{+0.0046}$ & $0.658_{-0.027}^{+0.019}$ & $\ldots$ & $\ldots$ \\ 
\hline
$\Lambda$CDM\ & DES Y1 + Planck (No Lensing) & $0.297_{-0.012}^{+0.016}$ & $0.795_{-0.013}^{+0.020}$ & $0.972_{-0.004}^{+0.006}$ & $0.0477_{-0.0012}^{+0.0016}$ & $0.686_{-0.014}^{+0.009}$ & $<0.47$ & $\ldots$ \\ 
\hline
$\Lambda$CDM\ & DES Y1 + JLA + BAO & $0.295_{-0.014}^{+0.018}$ & $0.768_{-0.023}^{+0.018}$ & $1.044_{-0.087}^{+0.019}$ & $0.0516_{-0.0080}^{+0.0050}$ & $0.672_{-0.034}^{+0.049}$ & $\ldots$ & $\ldots$ \\ 
\hline
$\Lambda$CDM\ & Planck + JLA + BAO & $0.306_{-0.007}^{+0.007}$ & $0.815_{-0.013}^{+0.015}$ & $0.969_{-0.005}^{+0.004}$ & $0.0483_{-0.0006}^{+0.0008}$ & $0.678_{-0.005}^{+0.007}$ & $<0.22$ & $\ldots$ \\ 
\hline
$\Lambda$CDM\ & \parbox{3cm}{ \strut DES Y1 + \\ Planck + JLA + BAO\strut } & $0.298_{-0.007}^{+0.007}$ & $0.802_{-0.012}^{+0.012}$ & $0.973_{-0.004}^{+0.005}$ & $0.0479_{-0.0008}^{+0.0007}$ & $0.685_{-0.007}^{+0.005}$ & $<0.26$ & $\ldots$ \\ 
\hline
\hline
$w$CDM\ & DES Y1 $\xi_\pm(\theta)$ & $0.274_{-0.042}^{+0.073}$ & $0.777_{-0.038}^{+0.036}$ & $\ldots$ & $\ldots$ & $\ldots$ & $\ldots$ & $-0.99_{-0.39}^{+0.33}$ \\ 
\hline
$w$CDM\ & DES Y1 $w(\theta)+\gamma_t$ & $0.310_{-0.036}^{+0.049}$ & $0.785_{-0.072}^{+0.040}$ & $\ldots$ & $\ldots$ & $\ldots$ & $\ldots$ & $-0.79_{-0.39}^{+0.22}$ \\ 
\hline
$w$CDM\ & DES Y1 3x2 & $0.284_{-0.030}^{+0.033}$ & $0.782_{-0.024}^{+0.036}$ & $\ldots$ & $\ldots$ & $\ldots$ & $\ldots$ & $-0.82_{-0.20}^{+0.21}$ \\ 
\hline
$w$CDM\ & Planck (No Lensing) & $0.222_{-0.024}^{+0.069}$ & $0.810_{-0.036}^{+0.029}$ & $0.960_{-0.007}^{+0.005}$ & $0.0334_{-0.0032}^{+0.0099}$ & $0.801_{-0.097}^{+0.045}$ & $\ldots$ & $-1.47_{-0.22}^{+0.31}$ \\ 
\hline
$w$CDM\ & DES Y1 + Planck (No Lensing) & $0.233_{-0.033}^{+0.025}$ & $0.775_{-0.021}^{+0.021}$ & $0.971_{-0.006}^{+0.004}$ & $0.0355_{-0.0039}^{+0.0050}$ & $0.775_{-0.040}^{+0.056}$ & $<0.65$ & $-1.35_{-0.17}^{+0.16}$ \\ 
\hline
$w$CDM\ & Planck + JLA + BAO & $0.303_{-0.008}^{+0.010}$ & $0.816_{-0.013}^{+0.014}$ & $0.968_{-0.006}^{+0.004}$ & $0.0479_{-0.0014}^{+0.0016}$ & $0.679_{-0.008}^{+0.013}$ & $<0.27$& $-1.02_{-0.05}^{+0.05}$ \\ 
\hline
$w$CDM\ & \parbox{3cm}{ \strut DES Y1 + \\ Planck + JLA + BAO\strut } & $0.301_{-0.010}^{+0.007}$ & $0.801_{-0.012}^{+0.011}$ & $0.974_{-0.005}^{+0.005}$ & $0.0483_{-0.0016}^{+0.0014}$ & $0.680_{-0.008}^{+0.013}$ & $<0.31$ & $-1.00_{-0.04}^{+0.05}$ \\ 
\hline
\end{tabular}
\end{table*}

\Svwide{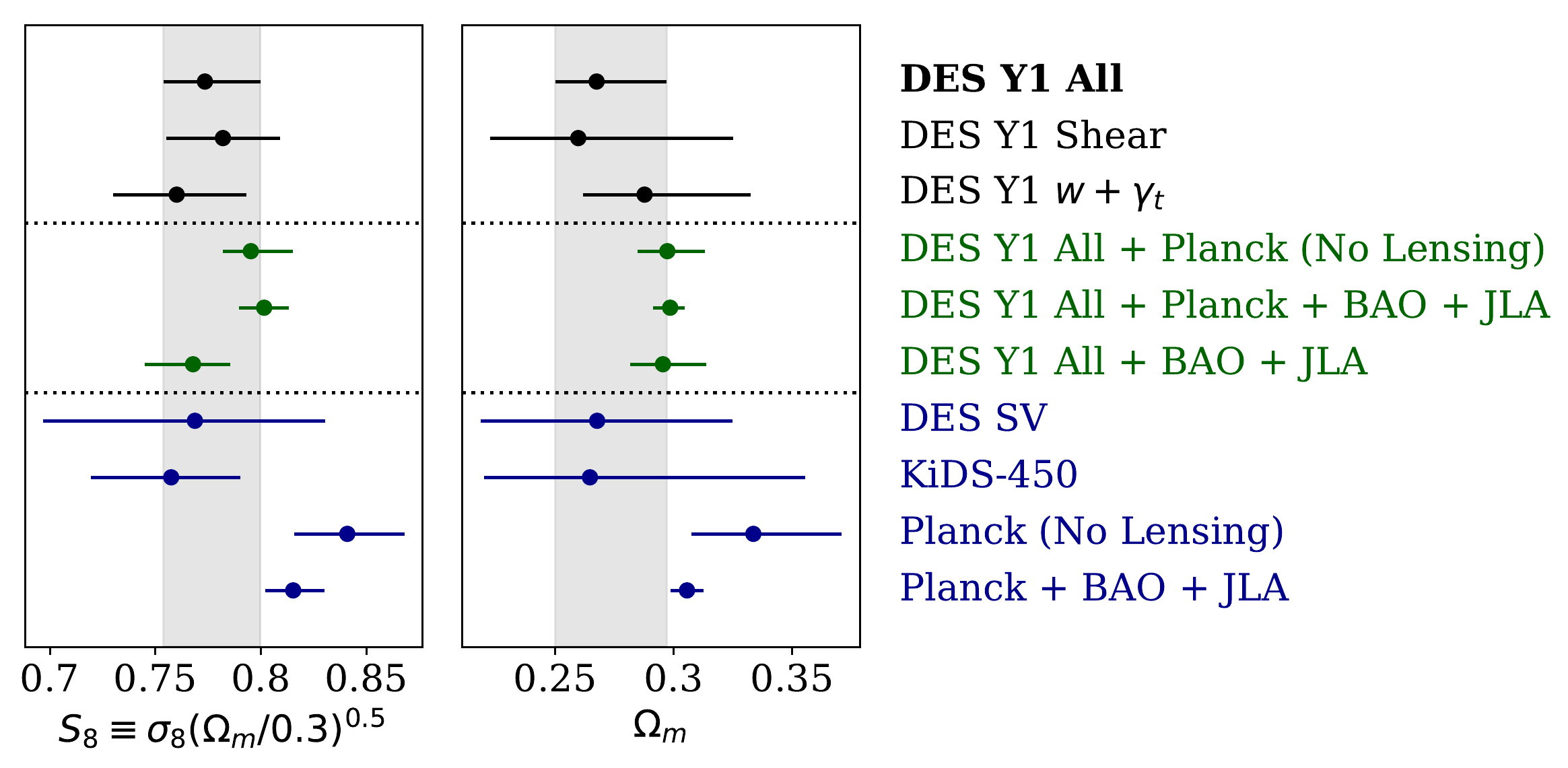}{68\% confidence levels for \lcdm\ on $S_8$ and $\Omega_m$
  from DES Y1 (different subsets considered in the top group, black); DES Y1
  with all three probes combined with other experiments (middle group, green);
  and results from previous experiments (bottom group, purple). Note that
  neutrino mass has been varied so, e.g., results shown for KiDS-450 were
  obtained by re-analyzing their data with the neutrino mass left free. The
  table includes only data sets that are publicly available so that we could
  re-analyze those using the same assumptions (e.g., free neutrino mass) as
  are used in our analysis of DES Y1 data.}

The results shown in Figure~\rf{ss8}, along with previous analyses such as
that using KiDS + GAMA data \cite{vanUitert:2017ieu}, are an important step
forward in the capability of combined probes from optical surveys to constrain
cosmological parameters. These combined constraints transform what has, for
the past decade, been a one-dimensional constraint on $S_8$ (which appears
banana-shaped in the $\Omega_m-\sigma_8$ plane) into tight constraints on both
of these important cosmological parameters. Figure~\rf{table_lcdm} shows the
DES Y1 constraints on $S_8$ and $\Omega_m$ along with some previous results
and in combination with external data sets, as will be discussed below. The
sizes of these parameter error bars from the combined DES Y1 probes are
comparable to those from the CMB obtained by Planck.

In addition to the cosmological parameters, these probes constrain important
astrophysical parameters. The intrinsic alignment (IA) signal is modeled to
scale as $A_{\rm IA}(1+z)^\salpha$; while the data do not constrain the power
law well ($\eta_{\rm IA}=-0.7\pm 2.2$), they are sensitive to the amplitude of the signal:
  \be
A_{\rm IA} = 0.44^{+0.38}_{-0.28}\qquad (95\% \ {\rm CL})
.\ee
Further strengthening evidence from the recent combined probes analysis of
KiDS~\cite{vanUitert:2017ieu,Joudaki:2017}, this result is the strongest
evidence to date of IA in a broadly inclusive galaxy sample; previously,
significant IA measurements have come from selections of massive elliptical
galaxies, usually with spectroscopic redshifts (e.g.\ \cite{Singh:2015}). The
ability of DES data to produce such a result without spectroscopic redshifts
demonstrates the power of this combined analysis and emphasizes the importance
of modeling IA in the pursuit of accurate cosmology from weak lensing. We are
able to rule out $A_{\rm IA} = 0$ at 99.76\% CL with DES alone and at
99.90\% CL with the full combination of DES and external data sets.  The
mean value of $A_{\rm IA}$ is nearly the same when combining with external
data sets, suggesting that IA self-calibration has been
effective. Interestingly, the measured amplitude agrees well with a prediction
made by assuming that only red galaxies contribute to the IA signal, and then
extrapolating the IA amplitude measured from spectroscopic samples of luminous
galaxies using a realistic luminosity function and red galaxy fraction
\method. Our measurement extends the diversity of galaxies with evidence of
IA, allowing more precise predictions for the behavior of the expected IA
signal.

The biases of the \redMaGiC\ galaxy samples in the five lens bins are shown in
Figure~\rf{bias} along with the results with fixed cosmology obtained in
\ggl\ and \clustering. The biases are measured to be
$b_1=1.42^{+0.13}_{-0.08}$,
$b_2=1.65^{+0.08}_{-0.12}$,
$b_3=1.60^{+0.11}_{-0.08}$,
$b_4=1.92^{+0.14}_{-0.10}$,
$b_5=2.00^{+0.13}_{-0.14}$.
Even when
varying a full set of cosmological parameters (including $\sigma_8$, which is
quite degenerate with bias when using galaxy clustering only) and 15 other
nuisance parameters, the combined probes in DES Y1 therefore constrain bias at the ten
percent level.

\Sfig{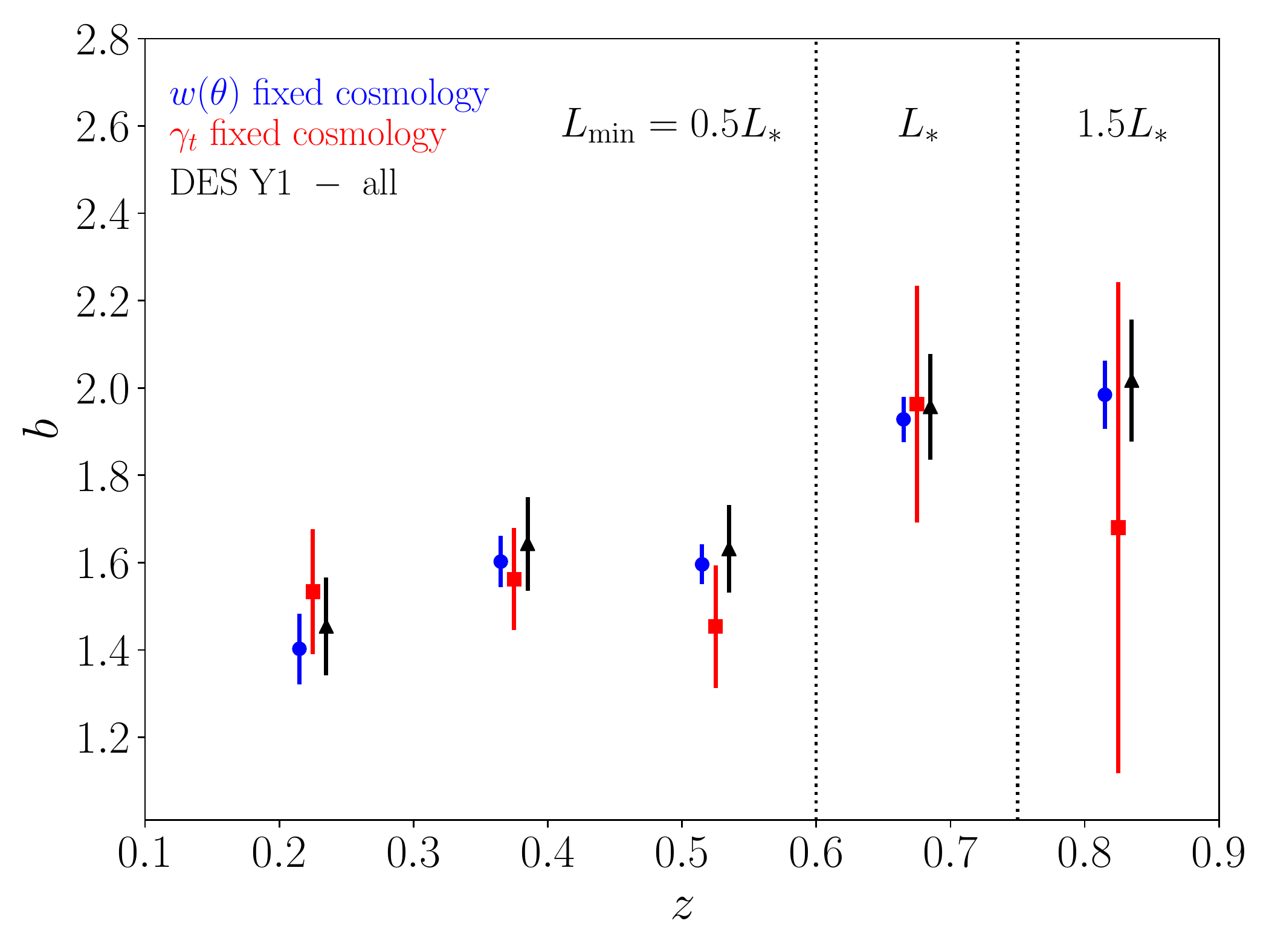}{The bias of the \redMaGiC\ galaxy samples in the five lens bins
  from three separate DES Y1 analyses. The two labeled ``fixed cosmology''
  use the galaxy angular correlation function $w(\theta)$ and galaxy--galaxy
  lensing $\gamma_t$ respectively, with cosmological parameters fixed at
  best-fit values from the 3x2 analysis, as described in \ggl\ and
  \clustering. The results labeled ``DES Y1 - all'' vary all 26 parameters
  while fitting to all three two-point functions.}

\subsection{\wcdm}

A variety of theoretical alternatives to the cosmological constant have been
proposed \cite{Frieman:2008sn}. For example, it could be that the cosmological
constant vanishes and that another degree of freedom, e.g., a very light
scalar field, is driving the current epoch of accelerated expansion.  Here we
restrict our analysis to the simplest class of phenomenological alternatives,
models in which the dark energy density is not constant, but rather evolves
over cosmic history with a constant equation of state parameter, $w$.  We
constrain $w$ by adding it as a seventh cosmological parameter. Here, too, DES
obtains interesting constraints on only a subset of the seven cosmological
parameters, so we show the constraints on the three-dimensional subspace
spanned by $\Omega_m$, $S_8$, and $w$. Figure~\rf{s_wg_d_w} shows the
constraints in this 3D space from cosmic shear and from galaxy--galaxy lensing
+ galaxy clustering. These two sets of probes agree with one another.  The
consistency in the three-dimensional subspace shown in Figure~\rf{s_wg_d_w},
along with the tests in the previous subsection, is sufficient to combine the
two sets of probes. The Bayes factor in this case was equal to 1878.
The combined constraint from all three two-point functions is also shown in
Figure~\rf{s_wg_d_w}.

\Swide{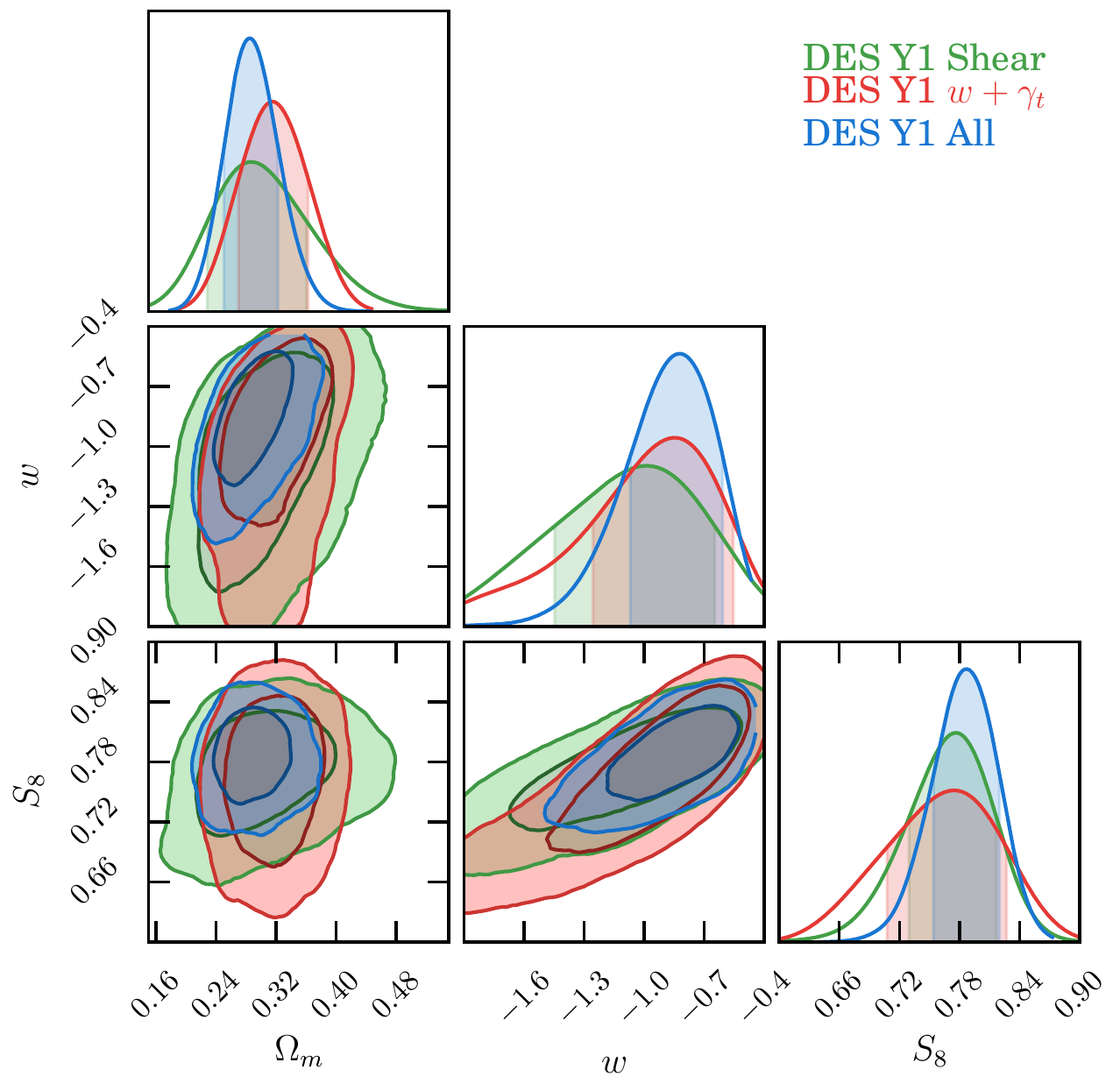}{Constraints on the three cosmological parameters $\sigma_8$,
  $\Omega_m$, and $w$ in \wcdm\ from DES Y1 after marginalizing over four other
  cosmological parameters and ten (cosmic shear only) or 20 (other sets of
  probes) nuisance parameters. The constraints from cosmic shear only (green);
  $w(\theta)+\gamma_t(\theta)$ (red); and all three two-point functions (blue) are
  shown. Here and below, outlying panels show the marginalized 1D posteriors and the corresponding 68\% confidence regions.}

The marginalized 68\% CL constraints on $w$ and on the other two cosmological parameters tightly
constrained by DES, $S_8$ and $\Omega_M$, are shown in
Figure~\rf{table_wcdm} and given numerically in Table~\ref{tab:post}. In the next section, we revisit the question of how
consistent the DES Y1 results are with other experiments. The marginalized
constraint on $w$ from all three DES Y1 probes is
\be
w = -0.82^{+0.21}_{-0.20} \eql{wres}
.\ee
 \Swide{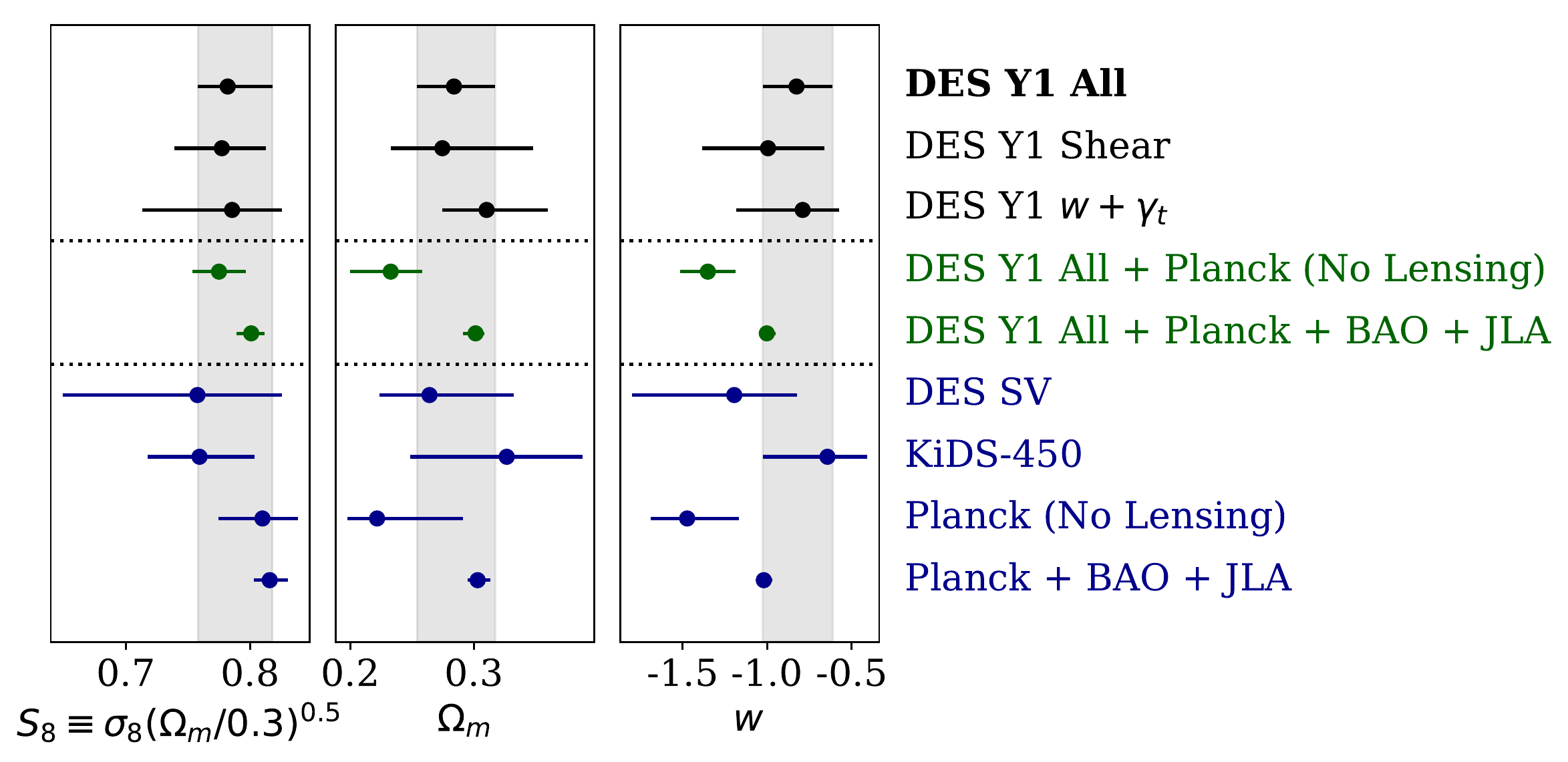}{68\% confidence levels on three cosmological parameters
   from the joint DES Y1 probes and other experiments for \wcdm.}

Finally, if one ignores any intuition or prejudice about the mechanism driving
cosmic acceleration, studying \wcdm\ translates into adding an additional
parameter to describe the data.  From a Bayesian point of view, the question
of whether \wcdm\ is more likely than \lcdm\ can again be addressed by
computing the Bayes factor.
Here the two models being compared are simpler: \lcdm\ and \wcdm. The Bayes
factor is
\be
R_w =  \frac{P( \vec D \vert w{\rm CDM} )}{P( \vec D \vert \Lambda{\rm CDM} )}\eql{ev2}
\ee
Values of $R_w$ less than unity would imply \lcdm\ is favored, while those greater than one argue that the introduction of the
additional parameter $w$ is warranted. The Bayes factor is $R_w = 0.39$ for
DES Y1, so although \lcdm\ is slightly favored, there is no compelling evidence to favor or disfavor an additional
parameter $w$.

It is important to note that, although our result in \ec{wres} is compatible with \lcdm, the most stringent test of the model  from DES Y1 is not this parameter, but rather the constraints on the parameters in the model shown in Figure~\rf{ss8} as compared with constraints on those parameters from the CMB measurements of the universe at high redshift. We turn next to that comparison.

\section{Comparison with external data}\label{sec:ext}

We next explore the cosmological implications of comparison and combination of
DES Y1 results with other experiments' constraints.  For the CMB, we take
constraints from Planck~\cite{Ade:2015xua}.  In the first subsection below, we
use only the temperature and polarization auto- and cross-spectra from Planck,
omitting the information due to lensing of the CMB that is contained in the
four-point function. The latter depends on structure and distances at late
times, and we wish in this subsection to segregate late-time information from
early-Universe observables. We use the joint TT, EE, BB and TE likelihood for
multipoles $\ell$ between 2 and 29 and the TT likelihood for $\ell$ between 30
and 2508 (commonly referred to as TT+lowP), provided by
Planck.\footnote{Late-universe lensing does smooth the CMB power spectra
  slightly, so these data sets are not {\it completely} independent of low
  redshift information.} In all cases that we have checked, use of WMAP
\cite{Bennett:2012zja} data yields constraints consistent with, but weaker
than, those obtained with Planck.  Recent results from the South Pole
Telescope ~\cite{Henning:2017nuy} favor a value of $\sigma_8$ that is
2.6-$\sigma$ lower than Planck, but we have not yet tried to incorporate these
results.

We use measured angular diameter distances
from the Baryon Acoustic Oscillation (BAO) feature
by the 6dF Galaxy Survey~\citep{Beutler:2011hx},
the SDSS Data Release 7 Main Galaxy Sample~\citep{Ross:2014qpa}, and BOSS Data
Release 12~\citep{Alam:2016hwk}, in each case extracting only the BAO
constraints. These BAO distances are all measured relative to the physical BAO
scale corresponding to the sound horizon distance $r_{\rm d}$; therefore, dependence
of $r_{\rm d}$ on cosmological parameters
must be included 
when determining the likelihood of any cosmological model (see
\cite{Alam:2016hwk} for details). 
We also use measures of luminosity distances from observations of distant Type
Ia supernovae (SNe) via the Joint Lightcurve Analysis (JLA) data from
\cite{Betoule:2014frx}.  

This set of BAO and SNe experiments has been
shown to be consistent with the \lcdm\ and \wcdm\ constraints from the CMB
\cite{Hinshaw:2012aka,Ade:2015xua}, so we can therefore sensibly merge this
suite of experiments---BAO, SNe, and Planck---with the DES Y1 results to
obtain unprecedented precision on the cosmological parameters.  We do not
include information about direct measurements of the Hubble constant because
those are in tension with this bundle of experiments \cite{Bernal:2016gxb}. 

\subsection{High redshift vs. low redshift in \lcdm}

The CMB measures the state of the Universe when it was 380,000 years old,
while DES measures the matter distribution in the Universe roughly ten billion
years later. Therefore, one obvious question that we can address is: Is the
\lcdm\ prediction for clustering today, with all cosmological parameters
determined by Planck, consistent with what DES observes?  This question, which
has of course been addressed by previous surveys~(e.g.,
\cite{Heymans:2013fya,Hildebrandt:2016iqg,vanUitert:2017ieu,Joudaki:2017}), is
so compelling because (i) of the vast differences in the epochs and conditions
measured; (ii) the predictions for the DES Y1 values of $S_8$ and $\Omega_m$
have no free parameters in \lcdm\ once the recombination-era parameters are
fixed; and (iii) those predictions for what DES should observe are very
precise, with $S_8$ and $\Omega_m$ determined by the CMB to within a few
percent. We saw above that $S_8$ and $\Omega_m$ are constrained by DES Y1 at
the few-percent level, so the stage is set for the most stringent test yet of
\lcdm\ growth predictions.  Tension between these two sets of constraints
might imply the breakdown of \lcdm.

Figure~\rf{dpnl_l} compares the low-$z$ constraints for \lcdm\ from all three
DES Y1 probes with the $z=1100$ constraints from the Planck anisotropy
data. Note that the Planck contours are shifted slightly and widened
significantly from those in Figure~18 of \cite{Ade:2015xua}, because we are
marginalizing over the unknown sum of the neutrino masses. We have verified
that when the sum of the neutrino masses is fixed as \cite{Ade:2015xua}
assumed in their fiducial analysis, we recover the constraints shown in their
Figure~18.

\Sfig{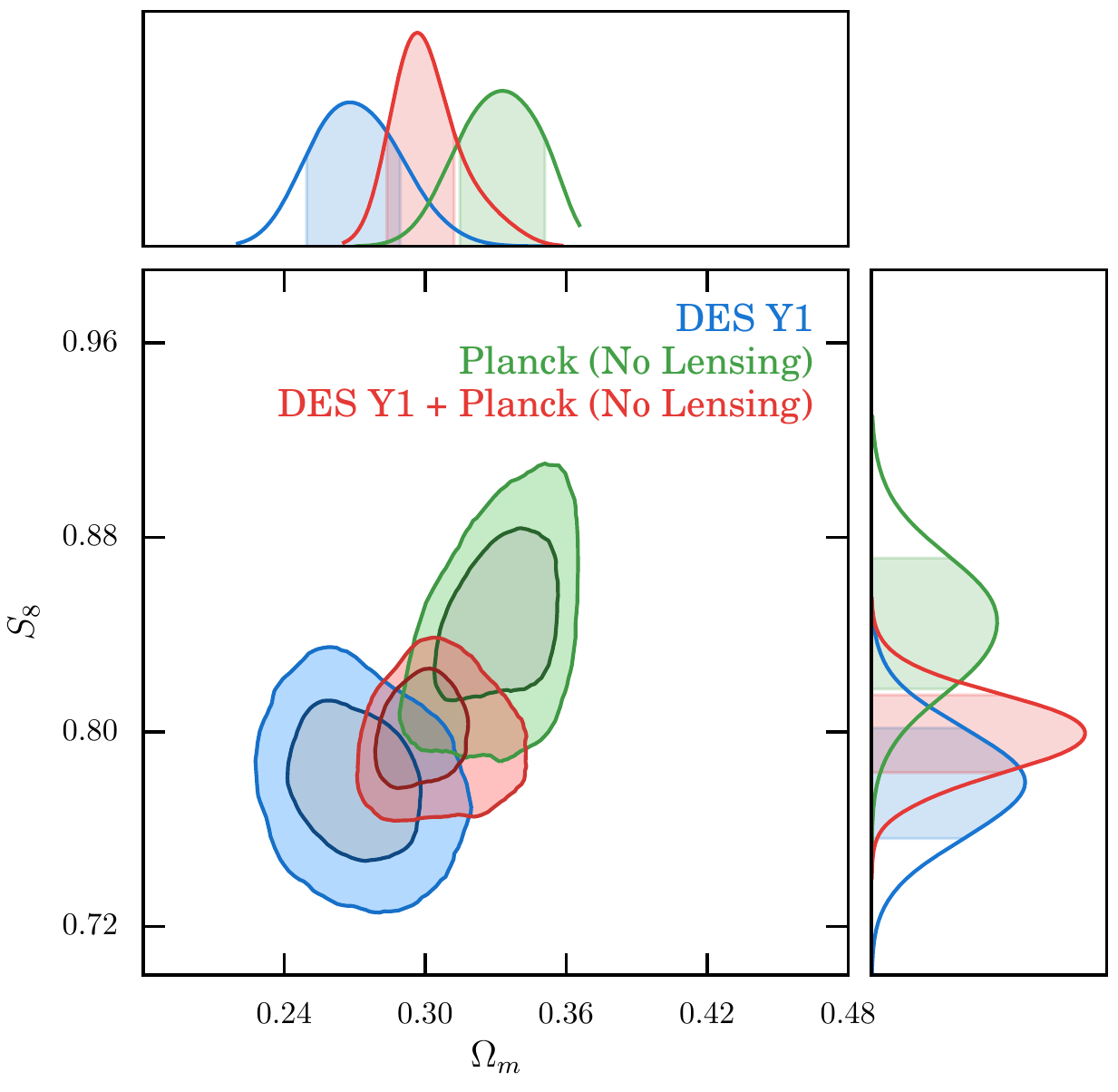}{\lcdm\ constraints from the three combined probes in DES Y1
  (blue), Planck with no lensing (green), and their combination (red). The
  agreement between DES and Planck can be quantified via the Bayes factor,
  which indicates that in the full, multi-dimensional parameter space, the two
  data sets are consistent (see text).}

The two-dimensional constraints shown in Figure~\rf{dpnl_l} visually hint at
tension between the Planck \lcdm\ prediction for RMS mass fluctuations and the
matter density of the present-day Universe and the direct determination by
DES.  The 1D marginal constraints differ by more than 1$\sigma$ in both $S_8$
and $\Omega_m$, as shown in Figure~\rf{table_lcdm}.  The KiDS
survey~\cite{Hildebrandt:2016iqg,Kohlinger:2017sxk,vanUitert:2017ieu,Joudaki:2017}
and, earlier, Canada-France Hawaii Telescope Lensing Survey (CHFTLenS;
  \cite{Heymans:2013fya,Fu:2014loa}) also report lower $S_8$ than Planck at
marginal significance.

However, a more quantitative measure of consistency in the full 26-parameter
space is the Bayes factor defined in \ec{evidence}.  As mentioned above, a
Bayes factor below 0.1 suggests strong inconsistency and one above 10 suggests
strong evidence for consistency.  The Bayes factor for combining DES and
  Planck (no lensing) in the $\Lambda$CDM model is $R=6.6$,
indicating ``substantial'' evidence for consistency on the Jeffreys scale, so
any inconsistency apparent in Figure~\rf{dpnl_l} is not statistically
significant according to this metric. In order to test the sensitivity of this
conclusion to the priors used in our analysis, we halve the width of the prior
ranges on all cosmological parameters (the parameters in the first section of
Table~\ref{tab:params}). For this case we find $R=0.75$;  despite
  dropping by nearly a factor of 10, $R$ it is still above 0.1 and therefore we are
  still passing the consistency test. The Bayes factor in \ec{evidence} compares
the hypothesis that two datasets can be fit by the same set of $N$ model
parameters (the null hypothesis), to the hypothesis that they are each allowed
an independent set of the $N$ model parameters (the alternative
hypothesis). The alternative hypothesis is naturally penalized in the Bayes
factor since the model requires an extra $N$ parameters. We also test an
alternative hypothesis where only $\Omega_m$ and $A_s$ are allowed to be
constrained independently by the two datasets; in this case we are introducing
only two extra parameters with respect to the null hypothesis. For this case,
we find $R=0.47$, which again indicates that there is no evidence for
inconsistency between the datasets.

We therefore combine the two data sets, resulting in the red contours in
Figure~\rf{dpnl_l}. This quantitative conclusion that the high-- and low--
redshift data sets are consistent can even be gleaned by viewing
Figure~~\rf{dpnl_l} in a slightly different way: if the true parameters lie
within the red contours, it is not unlikely for two independent experiments to
return the blue and green contour regions.

\Sfig{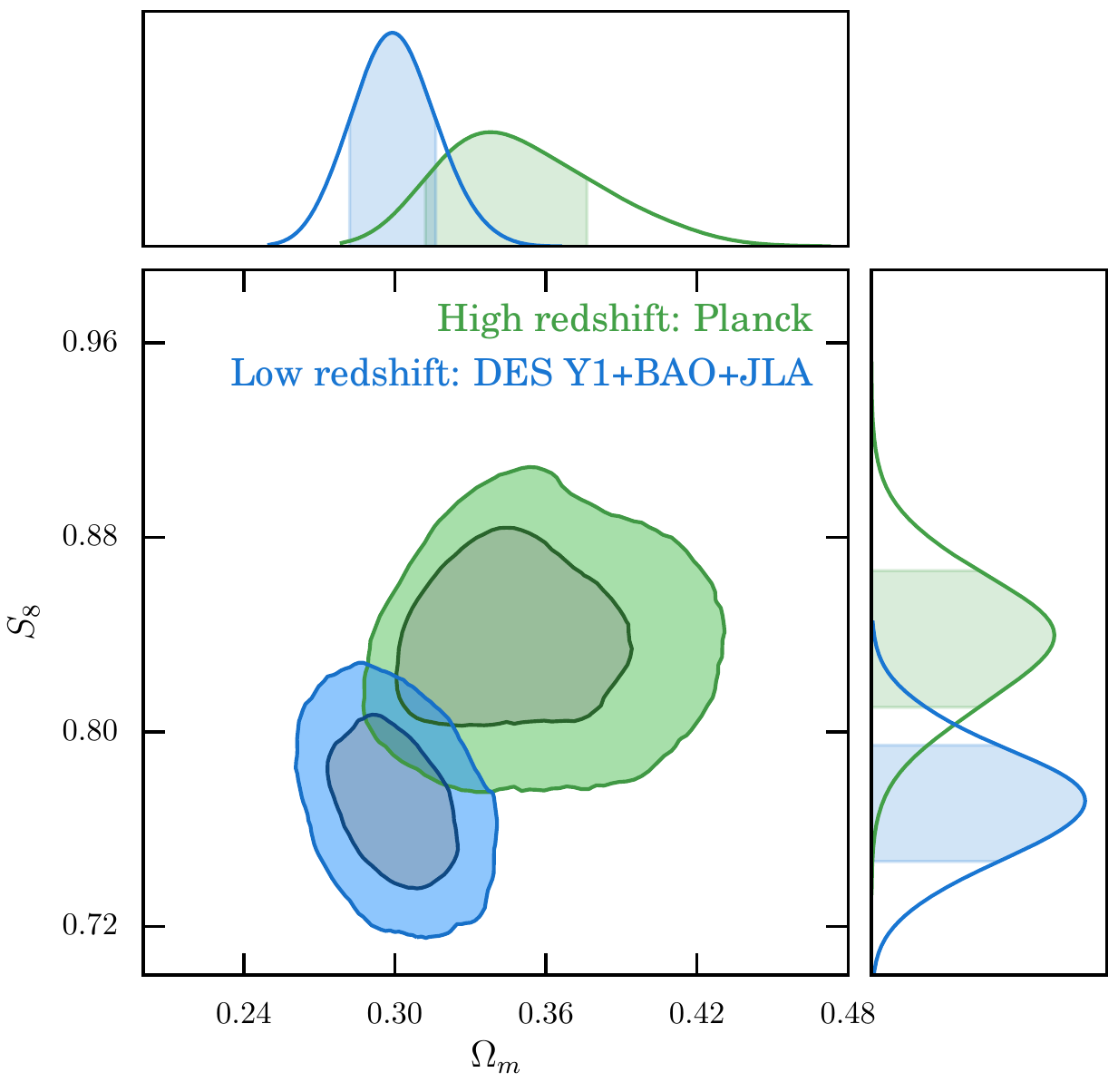}{\lcdm\ constraints from high redshift (Planck, without lensing)
  and multiple low redshift experiments (DES Y1+BAO+JLA), see text for
  references.}

Figure~\rf{hilo} takes the high-$z$ vs.~low-$z$ comparison a step further by
combining DES Y1 with results from BAO experiments and Type Ia supernovae. While these
even tighter low-redshift constraints continue to
favor slightly lower values of $\Omega_m$ and $S_8$ than Planck, the Bayes
factor
is 0.6, which neither favors nor disfavors the hypothesis that the two
sets of data, DES Y1+BAO+JLA on one hand and Planck on the other, are
described by the same set of cosmological parameters.

\Sfig{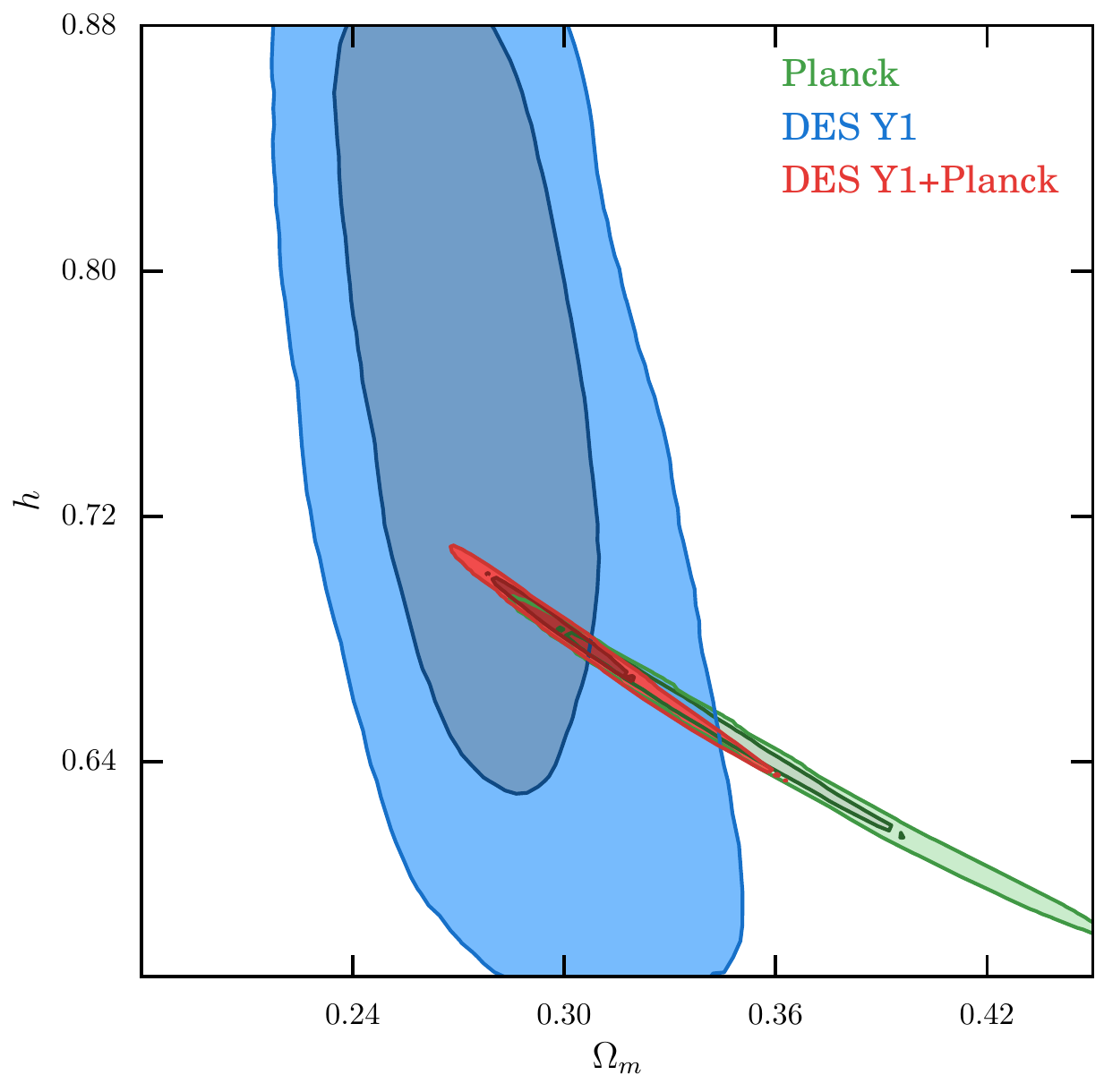}{\lcdm\ constraints from Planck with no lensing (green), DES Y1
  (blue) and the two combined (red) in the $\Omega_m,h$ plane. The positions
  of the acoustic peaks in the CMB constrain $\Omega_mh^3$ extremely well, and
  the DES determination of $\Omega_m$ breaks the degeneracy, leading to a
  larger value of $h$ than inferred from Planck only (see
  Table~\ref{tab:post}).}

The goal of this subsection is to test the \lcdm\ prediction for clustering in
DES, so we defer the issue of parameter determination to the next
subsections. However, there is one aspect of the CMB measurements combined
with DES that is worth mentioning here. DES data do not constrain the Hubble
constant directly. However, as shown in Figure~\rf{hubble}, the DES
\lcdm\ constraint on $\Omega_m$ combined with Planck's measurement of
$\Omega_mh^3$ leads to a shift in the inference of the Hubble constant (in the
direction of local measurements~\cite{Riess:2016jrr}). Since $\Omega_m$ is
lower in DES, the inferred value of $h$ moves up. As shown in the figure and
quantitatively in Table~\ref{tab:post}, the shift is greater than
$1\sigma$. As shown in Table~\ref{tab:post}, this shift in the value of $h$
persists as more data sets are added in.

\subsection{Cosmological Parameters in \lcdm}

To obtain the most stringent cosmological constraints, we now compare DES Y1
with the bundle of BAO, Planck, and JLA that have been shown to be consistent
with one another \cite{Ade:2015xua}. Here ``Planck'' includes the data from
the four-point function of the CMB, which captures the effect of lensing due
to large-scale structure at late times.  Figure~\rf{de_l} shows the
constraints in the $\Omega_m$--$S_8$ plane from this bundle of data sets and
from DES Y1, in the \lcdm\ model. Here the apparent consistency of the data
sets is borne out by the Bayes factor for dataset consistency
(\ecalt{evidence}):
\be
\frac{P({\rm JLA+Planck+BAO+DES\ Y1})}{P({\rm JLA+BAO+Planck})P({\rm DES\ Y1})} = 35
.\ee

\Sfig{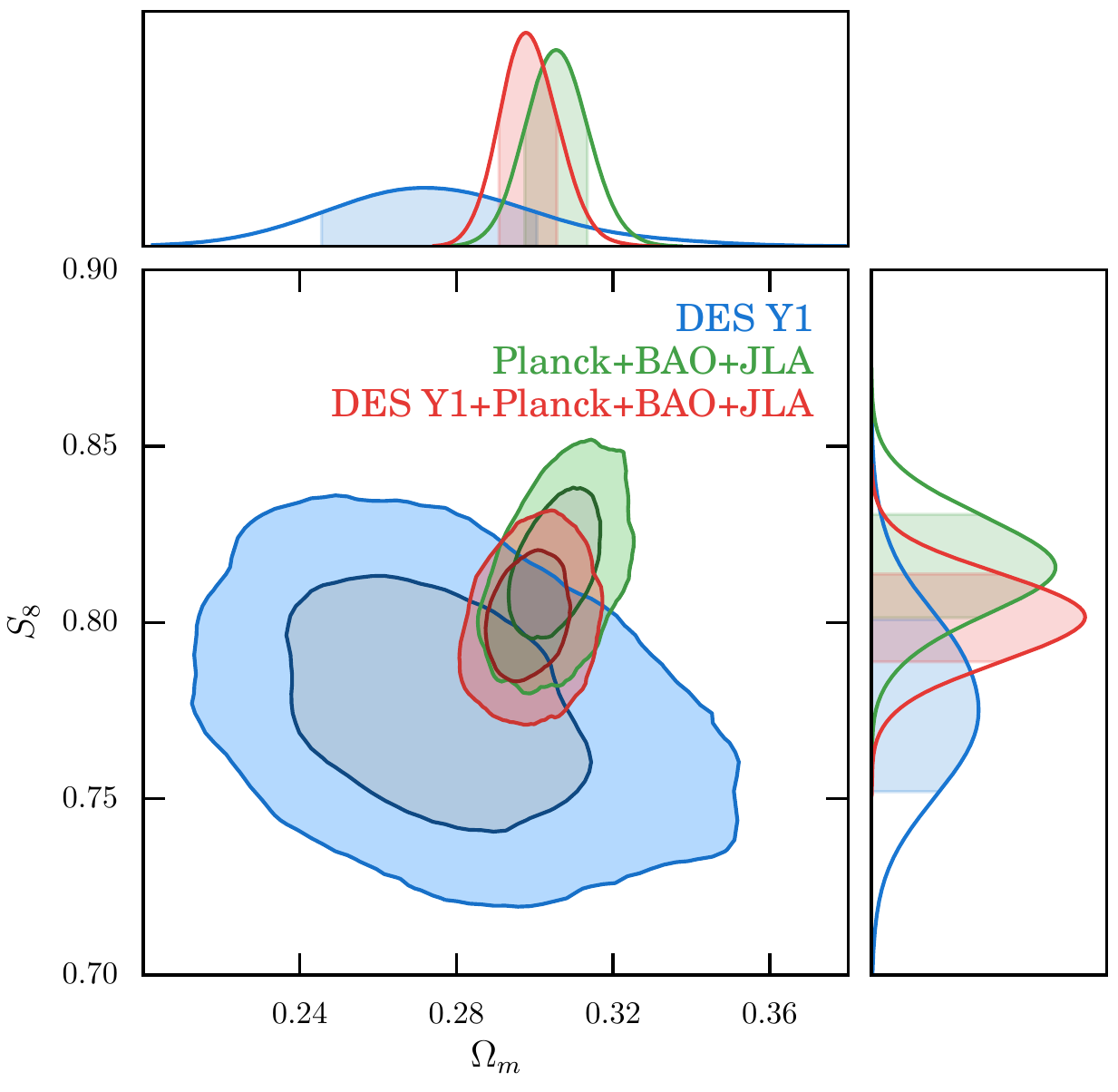}{\lcdm\ constraints from all three two-point functions within DES
  and BAO, JLA, and Planck (with lensing) in the $\Omega_m$-$S_8$ plane. }

Combining all of these leads to the tightest constraints yet on
\lcdm\ parameters, shown in Table ~\ref{tab:post}. Highlighting some of these:
at 68\% C.L., the combination of DES with these external data sets yields
\be
\Omega_m=0.298\pm 0.007\eql{omm}.
\ee
This value is about 1$\sigma$ lower than the value without DES Y1, with comparable error bars. The clustering amplitude is also constrained at the percent level:
\bea
\sigma_8&=&0.808^{+0.009}_{-0.017} \nonumber\\[0.2cm]
S_8&=&0.802\pm 0.012\eql{s8final}.
\eea
Note that fortuitously, because $\Omega_m$ is so close to $0.3$, the difference in the central values of $\sigma_8$ and $S_8$ is negligible. The combined result is about 1$\sigma$ lower than the inference without DES, and the constraints are tighter by about 20\%.

As mentioned above, the lower value of $\Omega_m$ leads to a higher value of the Hubble constant:
\bea
h &=& 0.658^{+0.019}_{-0.027}\qquad ({\rm Planck: No\ Lensing})
\vs
h &=& 0.685^{+0.005}_{-0.007}\qquad ({\rm DES\ Y1+JLA+BAO+Planck})\vs
\eea
with neutrino mass varied.

\subsection{\wcdm}
\Swide{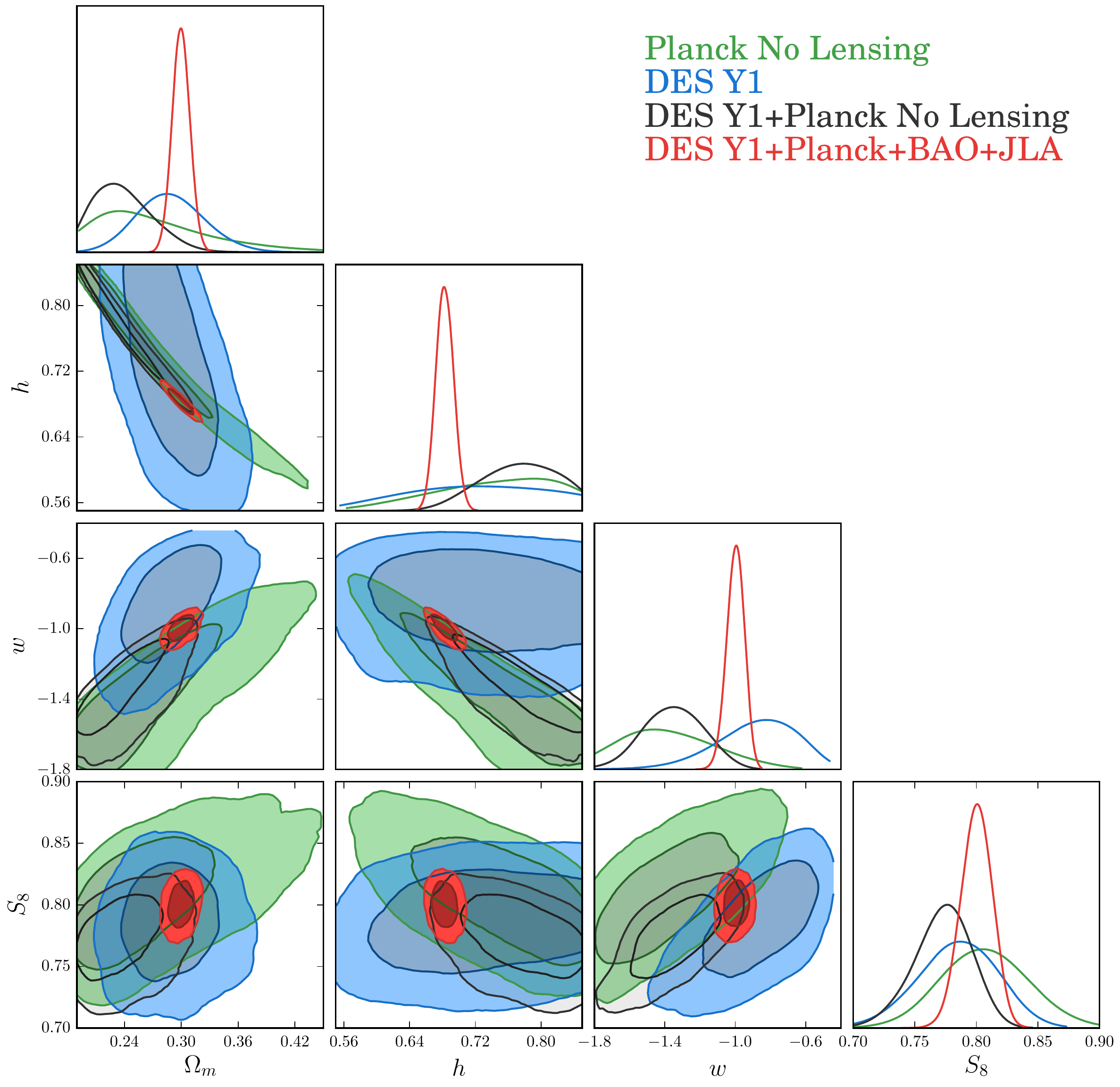}{\wcdm\ constraints from
  the three combined probes in DES Y1 and Planck with no lensing in the
  $\Omega_m$-$w$-$S_8$-$h$ subspace. Note the strong degeneracy between $h$
  and $w$ from Planck data. }
 
Figure~\rf{dpnl_w_4} shows the results in the extended \wcdm\ parameter space
using Planck alone, DES alone, the two combined, and the two with the addition
of BAO+SNe. As discussed in \cite{Ade:2015xua}, the constraints on the dark
energy equation of state from Planck alone are misleading. They stem from the
measurement of the distance to the last scattering surface, and that distance
(in a flat universe) depends upon the Hubble constant as well, so there is a
strong $w-h$ degeneracy. The low values of $w$ seen in Figure~\rf{dpnl_w_4}
from Planck alone correspond to very large values of $h$.
Since DES is not sensitive to the Hubble constant, it
does not break this degeneracy. Additionally, the Bayes factor in \ec{ev2}
that quantifies whether adding the extra parameter $w$ is warranted is
$R_w=0.7$. Therefore, opening up the dark energy equation of state is
not favored on a formal level for the DES+Planck combination.  Finally, the
Bayes factor for combining DES and Planck (no lensing) in \wcdm\ is equal to
10.3, indicating ``strong'' evidence that the two
  datasets are consistent. DES Y1 and Planck
  jointly constraint the equation of state to $w=-1.35_{-0.17}^{+0.16}$, which
  is about 2-sigma away from the cosmological-constant value. 

The addition of BAO, SNe, and Planck lensing data to the DES+Planck
combination yields the red contours in Figure~\rf{dpnl_w_4}, shifting the
solution substantially along the Planck degeneracy direction, demonstrating
(i) the problems mentioned above with the DES+Planck (no lensing) combination
and (ii) that these problems are resolved when other data sets are introduced
that restrict the Hubble parameter to reasonable values.  The Bayes factor for
combination of Planck (no lensing) with the low-$z$ suite of DES+BAO+SNe in
the \wcdm\ model is $R=89$ substantially more supportive of the
combination of experiments than the case for Planck and DES alone.  The
DES+Planck+BAO+SNe solution shows good consistency in the
$\Omega_m$--$w$--$S_8$ subspace and yields our final constraint on the dark
energy equation of state:
\be
w = -1.00_{-0.04}^{+0.05}\eql{combw}
.\ee
DES Y1 reduces the width of the allowed 68\% region by ten percent. The
evidence ratio $R_w=0.1$ for this full combination of data sets,
disfavoring the introduction of $w$ as a free parameter.

\subsection{Neutrino Mass}
\label{sec:neutrinos}

\Sfig{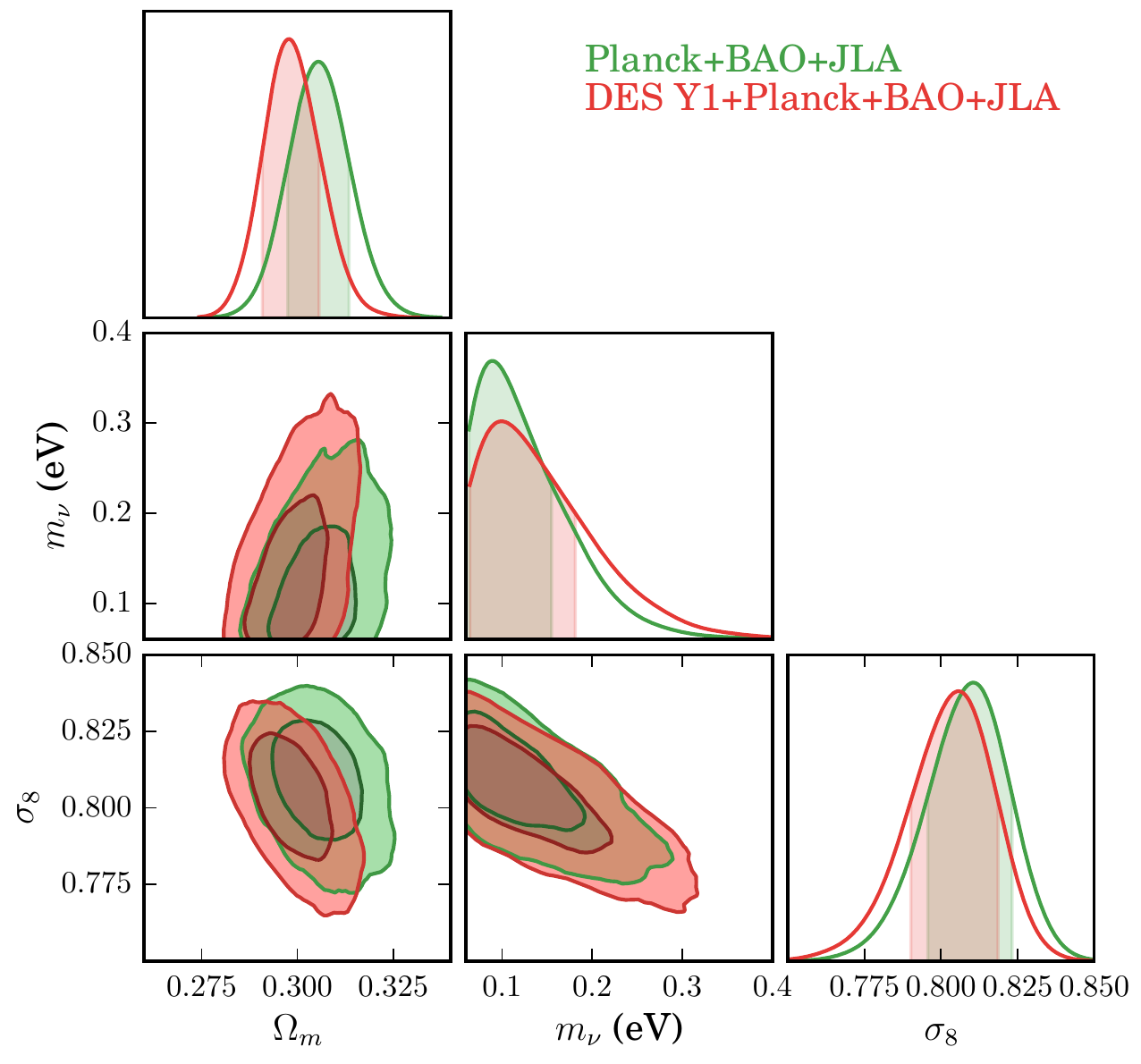}{\lcdm\ constraints on the sum of the neutrino masses from DES and
  other experiments. The lower power observed in DES can be accommodated
  either by lowering $\Omega_,$ or $\sigma_8$ or by increasing the sum of the
  neutrino masses.}

The lower power observed in DES (relative to Planck) has implications for the
constraint on the sum of the neutrino masses, as shown in
Figure~\rf{de_nu}. The current most stringent constraint comes from the cosmic
microwave background and Lyman-alpha forest~\cite{Yeche:2017upn}. The
experiments considered here (DES, JLA, BAO) represent an independent set so
offer an alternative method for measuring the clustering of matter as a
function of scale and redshift, which is one of the key drivers of the
neutrino constraints.  The 95\% C.L. upper limit on the sum of the neutrino
masses in \lcdm\ becomes less constraining:
\begin{equation}
  \sum m_\nu < 0.26\, {\rm  eV}\eql{mnu} .
\end{equation}
Adding in DES Y1 loosens the constraint by close to
20\% (from 0.22 eV). This is consistent with our finding that the
clustering amplitude in DES Y1 is slightly lower than expected in
\lcdm\ informed by Planck. The three ways of reducing the clustering amplitude
are to reduce $\Omega_m$, reduce $\sigma_8$, or increase the sum of the
neutrino masses. The best fit cosmology moves all three of these parameters
slightly in the direction of less clustering in the present day Universe.

 \Sfig{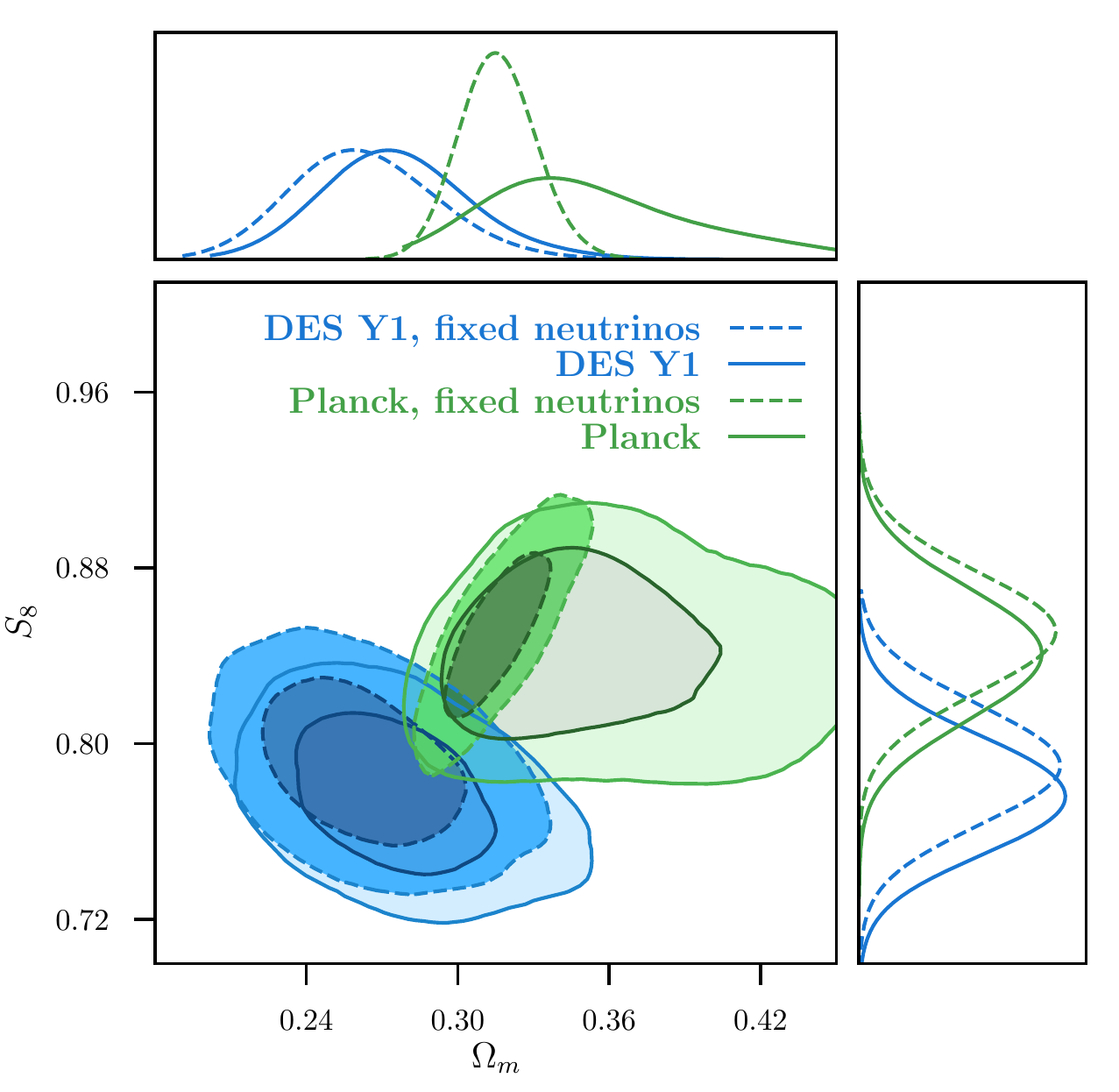}{\lcdm\ constraints on $\Omega_m$ and $\sigma_8$ from
   Planck without lensing and all three probes in DES. In contrast to all
   other plots in this paper, the dark contours here show the results when
   the sum of the neutrino masses was held fixed at its minimum allowed value
   of 0.06 eV.}
 
We may, conversely, be concerned about the effect of priors on $\Omega_\nu
h^2$ on the cosmological inferences in this paper.  The results for DES Y1 and
Planck depicted in Figure~\rf{dpnl_l} in \lcdm\ were obtained when varying the
sum of the neutrino masses. Neutrinos have mass \cite{Fukuda:1998} and the sum
of the masses of the three light neutrinos is indeed unknown, so this
parameter does need to be varied. However, many previous analyses have either
set the sum to zero or to the minimum value allowed by oscillation experiments
($\sum m_\nu=0.06$ eV), so it is of interest to see if fixing neutrino mass
alters any of our conclusions.  In particular: does this alter the level of
agreement between low- and high-redshift probes in \lcdm?
Figure~\rf{dpnl_fixednu} shows the extreme case of fixing the neutrino masses
to the lowest value allowed by oscillation data: both the DES and Planck
constraints in the $\Omega_m-S_8$ plane change. The Planck contours shrink
toward the low-$\Omega_m$ side of their contours, while the DES constraints
shift slightly to lower $\Omega_m$ and higher $S_8$.  The Bayes factor for the
combination of DES and Planck in the \lcdm\ space changes from $R=6.6$ to
$R=3.4$ when the minimal neutrino mass is enforced. DES and Planck
  therefore continue to agree, as seen in Figure~\rf{dpnl_fixednu}: when the
  neutrino mass is fixed, the area in the $\Omega_m-S_8$ plane allowed by
  Planck is much smaller than when $\Omega_\nu h^2$ varies, but there remains
  a substantial overlap between the Planck and DES contours. 
 
Finally, fixing the neutrino mass allows us to compare directly to previous
analyses that did the same. Although there are other differences in the
analyses, such as the widths of the priors, treatments of systematics, and
covariance matrix generation, fixing the neutrino mass facilitates a more
accurate comparison. On the main parameter $S_8$ within \lcdm, again with
neutrino mass fixed, the comparison is:
\begin{eqnarray}
 S_8 &=& 0.793^{+0.019}_{-0.026}\qquad \text{DES\ Y1}
 \vs
 &=&0.801\pm0.032 \qquad \text{KiDS+GAMA \cite{vanUitert:2017ieu}}
 \vs
&=& 0.742\pm0.035 \qquad \text{KiDS+2dFLenS+BOSS \cite{Joudaki:2017}},\eql{s8nu}
\vs
\end{eqnarray}
so we agree with KiDS+GAMA, and differ from KiDS+2dFLenS+BOSS by only about 1.2$\sigma$, indicating good
  statistical agreement.

\section{Conclusions}\label{sec:con}

We have presented cosmological results from a combined analysis of galaxy
clustering and weak gravitational lensing, using imaging data from the first
year of DES.  These combined probes demonstrate that cosmic surveys using
clustering measurements have now attained constraining power comparable to the
cosmic microwave background in the $\Omega_m$--$S_8$ plane, heralding a new
era in cosmology. The combined constraints on several cosmological parameters
are the most precise to date.

The constraints on $\Omega_m$ from the CMB stem from the impact of the matter
density on the relative heights of the acoustic peaks in the cosmic plasma
when the universe was only 380,000 years old and from the distance between us
today and the CMB last scattering surface. The CMB constraints on $S_8$ are an
expression of both the very small RMS fluctuations in the density at that
early time and the model's prediction for how rapidly they would grow over
billions of years due to gravitational instability. The measurements
themselves are of course in microwave bands and probe the universe when it was
extremely smooth. DES is different in every way: it probes in optical bands
billions of years later when the universe had evolved to be highly
inhomogeneous. Instead of using the radiation as a tracer, DES uses galaxies
and shear. It is truly extraordinary that a simple model makes consistent
predictions for these vastly different sets of measurements.

The results presented here enable precise tests of the \lcdm\ and \wcdm\ models, as shown in Figures~\rf{dpnl_l} and \rf{dpnl_w_4}. Our main findings are:
\begin{itemize}
\item DES Y1 constraints on $\Omega_m$ and $S_8$ in \lcdm\ are competitive (in
  terms of their uncertainties) and compatible (according to tests of the
  Bayesian evidence) with constraints derived from Planck observations of the
  CMB. This is true even though the visual comparison (Figure~\rf{dpnl_l}) of
  DES Y1 and Planck shows differences at the 1 to 2-$\sigma$ level, in the
  direction of offsets that other recent lensing studies have reported.
\item The statistical consistency allows us to combine DES Y1 results with
  Planck, and, in addition, with BAO and supernova data sets. This yields $S_8
  = 0.802\pm 0.012$ and $\Omega_m = 0.298\pm 0.007$ in
  \lcdm, the tightest such constraints to date (Figure~\rf{de_l}).
\item The \wcdm\ likelihoods from DES and Planck each constrain $w$ poorly;
  moreover, allowing $w$ as a free parameter maintains the consistency of
    the two data sets. DES is also consistent with the bundle of
  Planck, BAO, and supernova data, and this combination tightly constrains the
  equation-of-state parameter, $w = -1.00_{-0.04}^{+0.05}$
  (Figure~\rf{dpnl_w_4}).
\item The two-point functions measured in DES Y1 contain some information on
  two other open questions in cosmological physics: the combination of DES and
  Planck shifts the Planck constraints on the Hubble constant by more than
  $1\sigma$ in the direction of local measurements (Figure~\rf{hubble}), and
  the joint constraints on neutrino mass slightly loosens the bound from
  external experiments to $\sum m_{\nu}<0.26\,$eV (95\% C.L.)
  (Figure~\rf{de_nu}).
\item All results are based on redundant implementations and tests of the most
  critical components. They are robust to a comprehensive set of checks that
  we defined a priori and made while blind to the resulting cosmological
  parameters (see Section \ref{sec:blind} and Appendix \ref{sec:unblind}). All
  related analyses, unless explicitly noted otherwise, marginalize over the
  relevant measurement systematics and neutrino mass.
\item Joint analyses of the three two-point functions of weak lensing and
  galaxy density fields have also been executed recently by the combination of
  the KiDS weak lensing data with the GAMA \cite{vanUitert:2017ieu} and
  2dfLenS \cite{Joudaki:2017} spectroscopic galaxy surveys, yielding
  \lcdm\ bounds on $S_8$ that are in statistical agreement with ours; see \ec{s8nu}. 
    DES Y1 uncertainties are roughly $\sqrt{2}$ narrower than
  those from KiDS-450; while one might have expected a greater improvement
  considering the $\sim$3$\times$ increase in survey area, we caution against
  any detailed comparison of values or uncertainties until the analyses are
  homogenized to similar choices of scales, priors on neutrino masses, and
  treatments of observational systematic uncertainties.
\end{itemize}

The next round of cosmological analyses of DES data will include data from the
first three years of the survey (DES Y3), which cover more than three times as
much area to greater depth than Y1, and will incorporate constraints from
clusters, supernovae, and cross-correlation with CMB lensing, shedding more
light on dark energy and cosmic acceleration.

\section*{Acknowledgments}
  
\input{ack.tex}

We are grateful to An\v{z}e Slosar for helpful communications. Many of the figures
in this paper were produced with {\tt chainconsumer}~\cite{Hinton2016}. This
research used resources of the National Energy Research Scientific Computing
Center, a DOE Office of Science User Facility supported by the Office of
Science of the U.S. Department of Energy under Contract No. DE-AC02-05CH11231.

\bibliography{refs,des_y1kp_short}
 
 \appendix

 \section{Unblinding Tests}\label{sec:unblind}
 
Here we describe some of the results of the tests enumerated in
\S\ref{sec:blind}.  The most relevant metrics are the values of the
cosmological parameters best constrained by DES Y1, namely $\Omega_m$ and
$S_8$.  We report here on the few instances in which the robustness tests
yielded shifts in either the values or the uncertainties on $S_8$ or
$\Omega_m$ exceeding 10\% of their 68\% CL intervals.

Fig,~\rf{cs_cl} shows the result of test \ref{cs_vs_cl}.  As {\tt CosmoSIS}
and {\tt CosmoLike} use the same data and models, there should in principle be
no difference between them except for the sampling noise of their finite MCMC
chains.  {\tt CosmoSIS} yields error bars on $\Omega_m$ slightly smaller than
those obtained from {\tt CosmoLike}, with $<0.2\sigma$ change in central
value.  The $S_8$ constraints agree to better than a percent and the error
bars to within 3\%. These numbers and the contours shown in Figure~\rf{cs_cl}
improved over the results obtained before unblinding, when the difference in
the error bars was larger. Longer {\tt emcee} chains account for the
improvement, so it is conceivable that these small differences --- which do
not affect our conclusions --- go away with even longer chains.
% cl: omega_m 0.282496991501 0.0340958960804
%S8 0.781169656812 0.0247893526873
%%% cs
%cosmological_parameters--omega_m 0.275996022162 0.0259777747379
%cs S8 0.779734103955 0.0221242898779

\Sfig{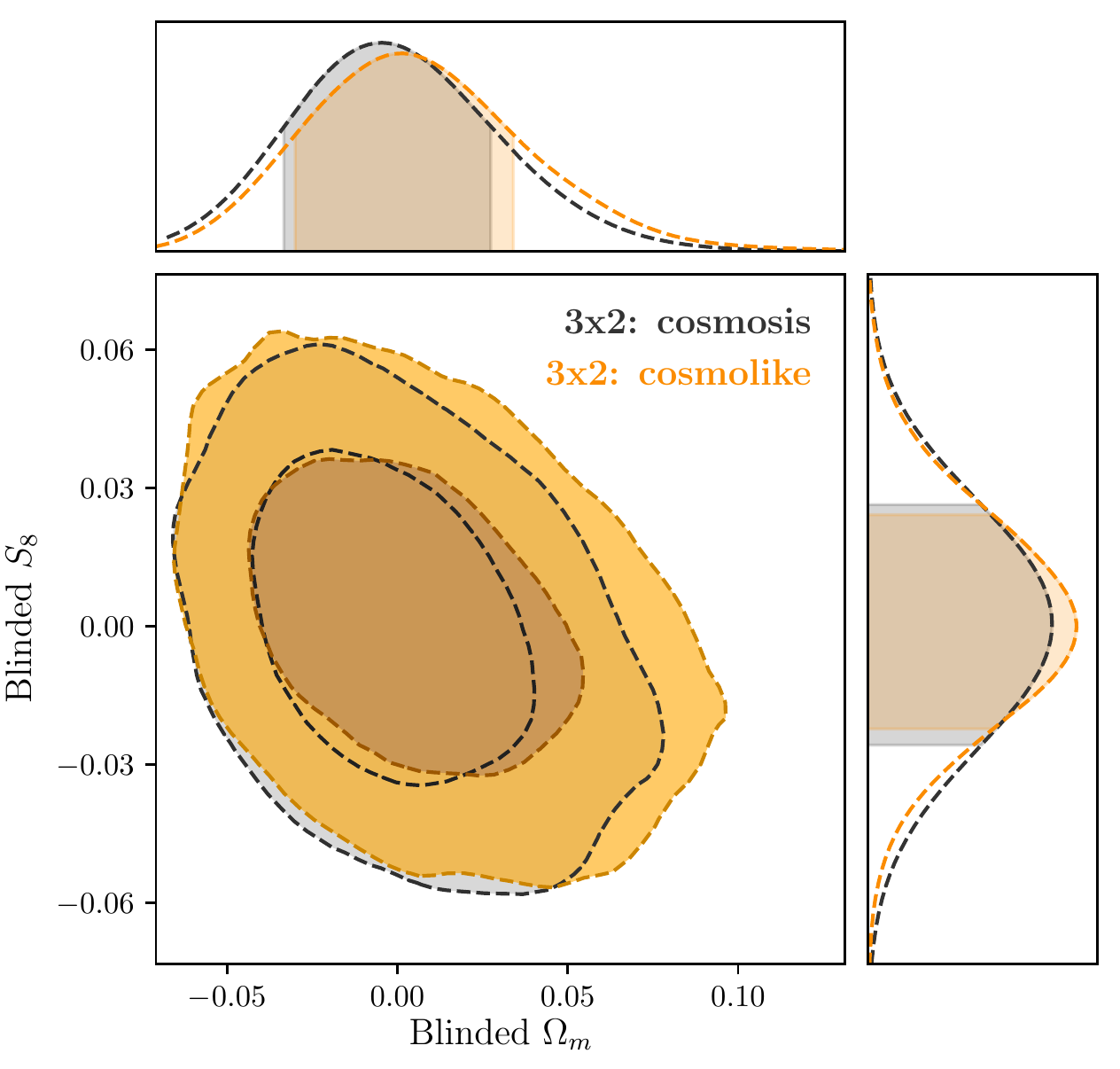}{Blinded constraints on $\Omega_m$ and $S_8$ from all three
  2-point functions in DES Y1 using two separate analysis pipelines on the
  data. Both contours are shifted by the means of the posteriors obtained from
  {\tt CosmoSIS}, so that the {\tt CosmoLike} contours could in principle be
  centered away from the origin. This figure was made prior to unblinding,
  thus without the update to the covariance described in
  Appendix~\ref{sec:covupdate}.}

When carrying out test~\ref{posteriors_vs_priors}, we found that for both
\metacal\ and \im, almost all of the parameters were tightly constrained to
lie well within their sampling ranges. The lone exception was the power law of
the intrinsic alignment signal, $\salpha$, which had an error that is large
relative to the prior, but this was entirely expected, as our simulations
indicated that the Y1 data have little constraining power on $\salpha$.  For
those parameters with more informative priors, the posteriors typically fell
close to the priors, indicating that the data were consistent with the
calibrations described in \shape\ and \photozpaper. One exception was the
\imshape\ value $\Delta z_s^4$, the shift in the mean value of the redshift in
the 4th source bin, where the posterior and prior differed by close to
1$\sigma$.

We next compare the \metacal\ and \imshape\ constraints in the
  $\Omega_m$-$S_*$ plane, noting that Figure 12 of \shearcorr\ already shows
  good agreement between the two pipelines on inferences purely with cosmic
  shear. Figure~\rf{di3} shows that when all $3\times2$-point data are
combined, \metacal\ and \imshape\ are in good
  agreement. Note also that their corresponding data vectors are not directly
comparable, since they bin and weight the source galaxies differently and thus
have distinct redshift distributions---they can be properly compared only in
cosmological-parameter tests such as this.

\Sfig{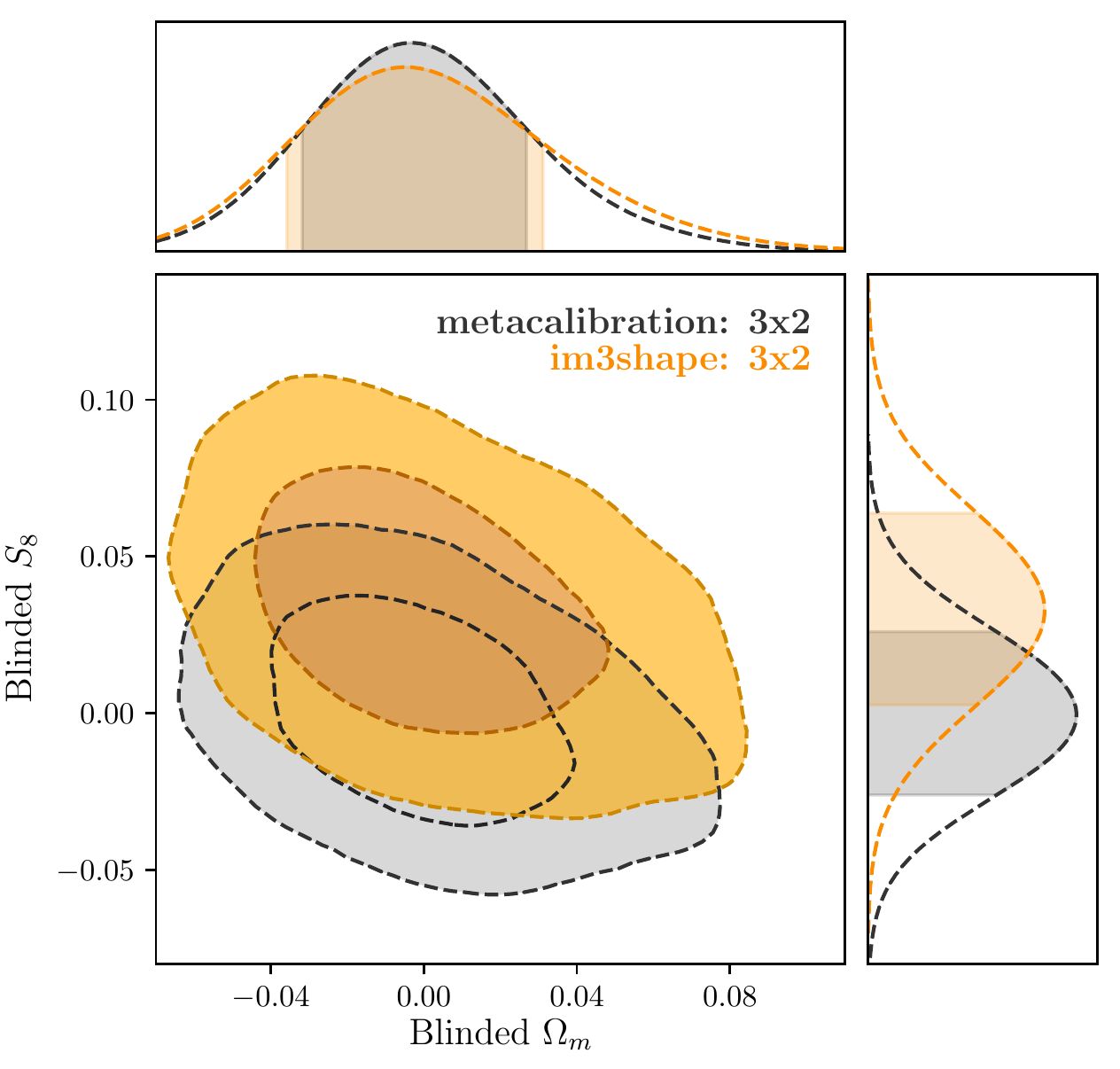}{Blinded constraints from DES Y1 on $\Omega_m$ and $S_8$ from all
  three combined probes, using the two independent shape pipelines
  \metacal\ and \im. }

For test \ref{fiducial_vs_largescale}, we deleted from the data vector 
angular scales
$<20$ arcmin from $\xi_+$, $<150$ arcmin from $\xi_-$,  $<65$ arcmin from $\gamma_t$, 
and $<50$ arcmin from $w(\theta)$.  The cosmological parameter constraints expanded slightly,
as expected, but shifted by much less than $1\sigma.$

Finally, although we looked at these blinded, Figure~\rf{table_1d_all} shows
the posteriors of all 20 nuisance parameters used to model the data. Note the
agreement of the two sets of probes with each other and with the priors on the
parameters.

\Sswide{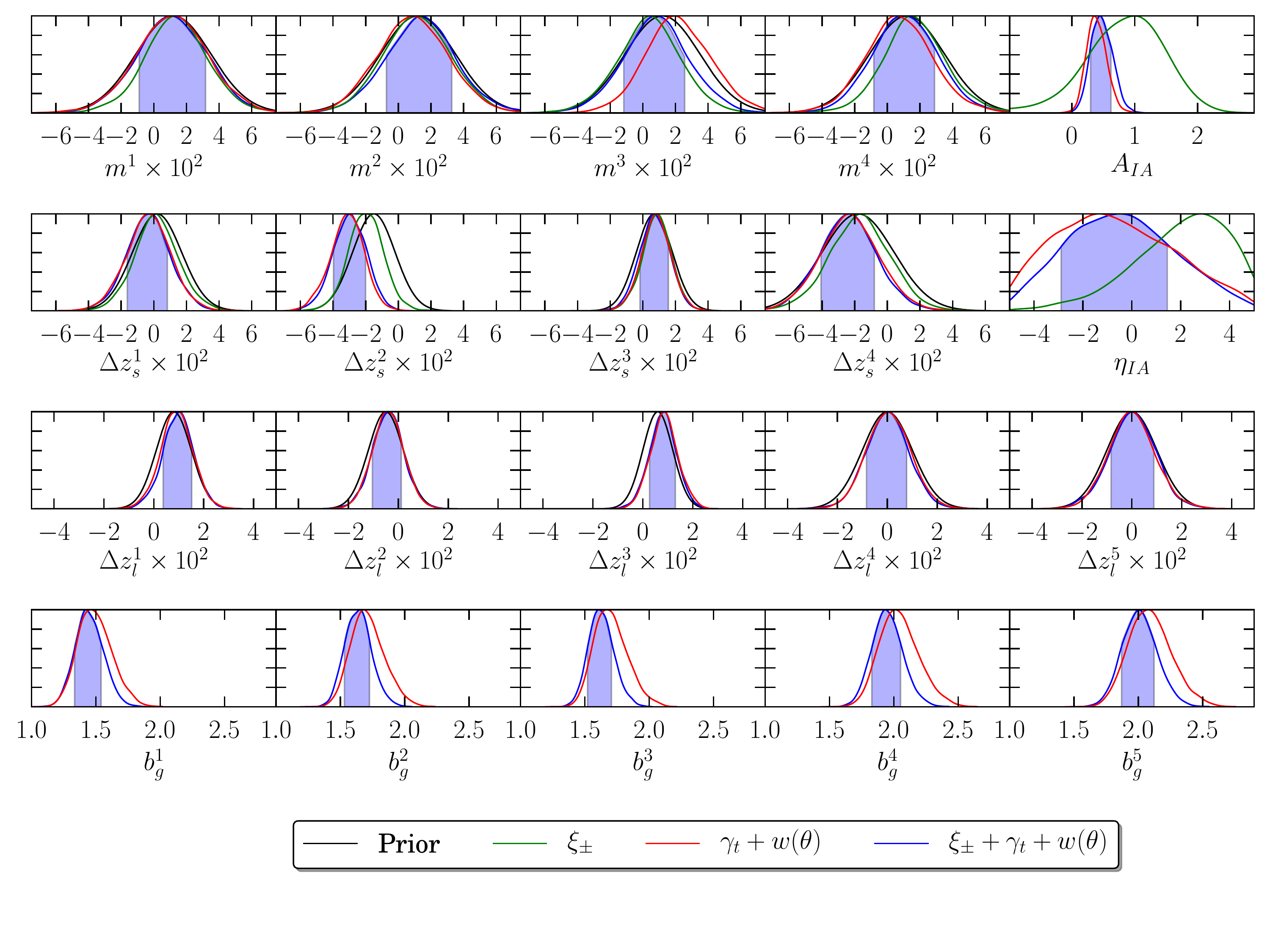}{The posteriors from cosmic shear; from
  $w(\theta)+\gamma_t(\theta)$; and for all three probes using the
  \metacal\ pipeline for all 20 nuisance parameters used in the
  \lcdm\ analysis.  The priors are also shown. There are no priors for the
  bias and intrinsic alignment parameters, the biases and the lens shifts are
  not constrained by $\xi_\pm$. Therefore, the bottom panels have only two
  curves: posteriors from $w(\theta)+\gamma_t(\theta)$ and from all three
  probes. Similarly, there are only three curves for the two intrinsic
  alignment parameters. }

Before unblinding, we listed several additional robustness tests that would be
carried out after unblinding.  These are described in
Appendix~\ref{sec:robustness}.

 \section{Robustness of Results}
 \label{sec:robustness}

Here we test the impact on the final results of some of the choices made during analysis. These tests, conducted while unblinded but identified beforehand, supplement those described in \S\ref{sec:blind}.
 
All of our inferences require assumptions about the redshift distributions for the source and lens galaxies. We have quantified the uncertainties in the redshift distributions with a shift parameter, as described in and around \ec{nidef}. This allows for the means of the distributions to change but does not allow for any flexibility in the shapes. We now check that the uncertainty in the photometric distributions in the source bins is adequately captured by using the BPZ redshift distribution accompanied by the free shift parameter in each bin. Instead of redshift distributions obtained via BPZ, we use those obtained directly from the COSMOS data, as described in \photozpaper. As shown in Figure 4 there, the shapes of the redshift distributions are quite different from one another, so if we obtain the same cosmological results using these different shape $n(z)$'s, we will have demonstrated that the detailed shapes do not drive the constraints. Again we allow for a free shift in each of the source distributions. Figure~\rf{dcosmos} shows that the ensuing constraints are virtually identical to those that use the BPZ $n(z)$'s for the source galaxies, suggesting that our results are indeed sensitive only to the means of the redshift distributions in each bin, and not to the detailed shapes.
 
 \Sfig{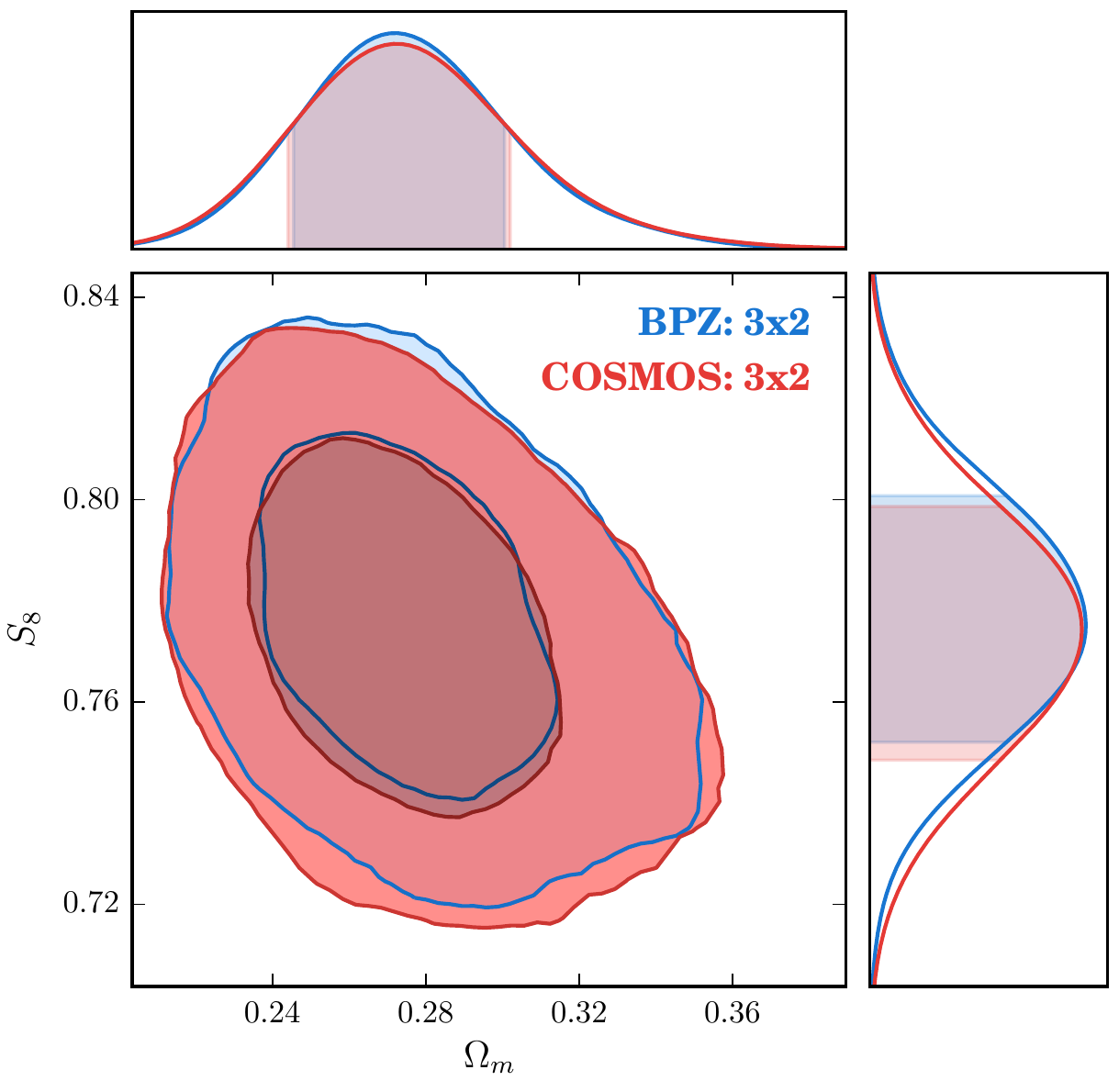}{Constraints on $\Omega_m$ and $S_8$ when using the
   shifted BPZ redshift distributions as the default for $n^i_s(z),$ compared with
   those obtained when using the COSMOS redshift distribution, which
   have different shape, as seen in Figure~4 of \cite{photoz}.}
 
 We also considered the impact of the choices made while computing the
 covariance matrix. These choices require assumptions about all 26 parameters
 that are varied. We generated an initial covariance matrix assuming fiducial
 values for these parameters, but then after unblinding, recomputed it using
 the means of the posteriors of all the parameters as input. How much did this
 (small) change in the covariance matrix affect our final results?
 Figure~\rf{dcov} shows that the updated covariance matrix had essentially no
 impact on our final parameter determination.
 
 \Sfig{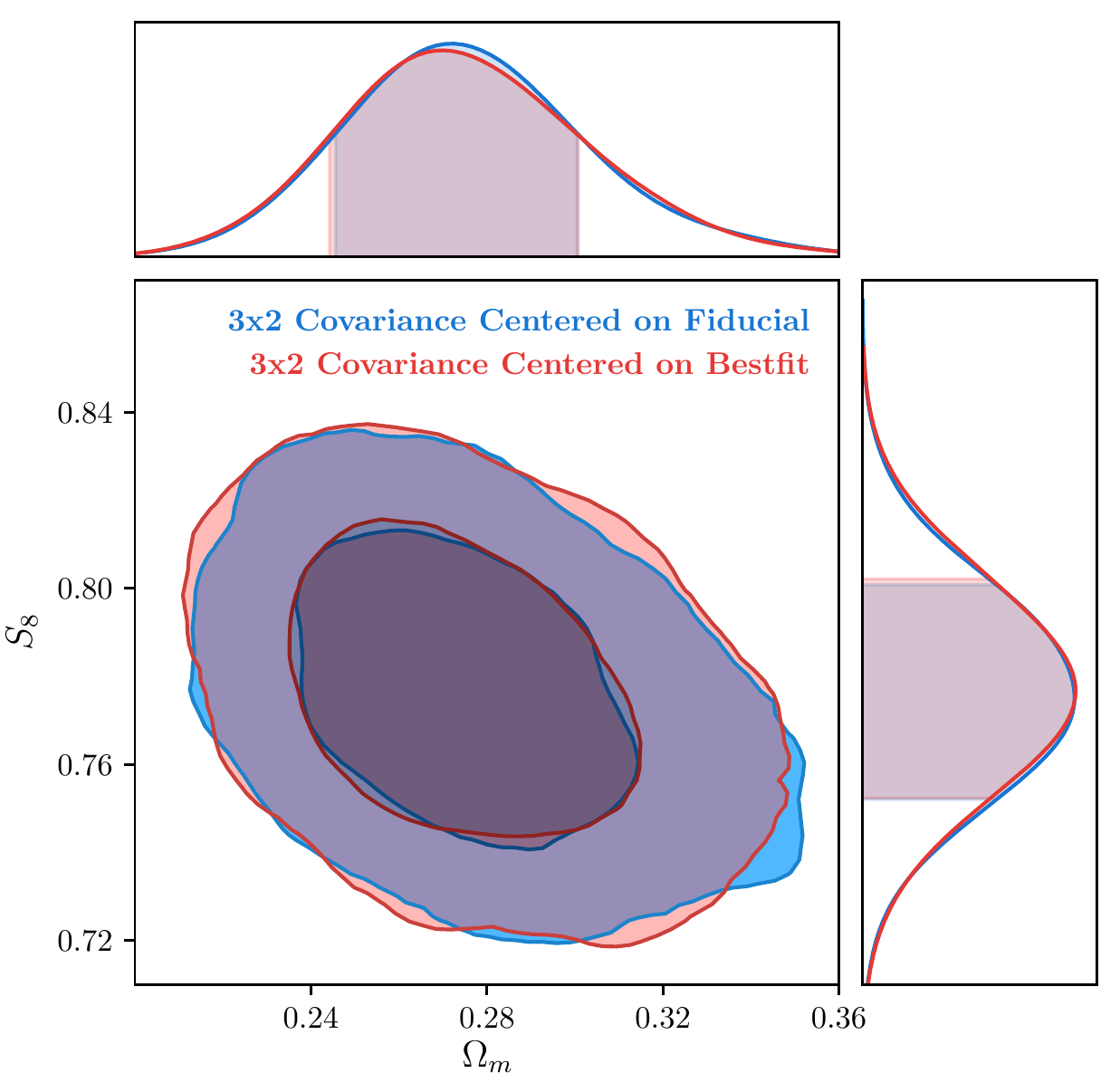}{Constraints on $\Omega_m$ and $S_8$ using the fiducial covariance
   matrix, and using the covariance based on the cosmological model centered
   on the means of the posteriors ('Bestfit') obtained after unblinding. The two agree
   very well, indicating little dependence on the fiducial model assumed for
   the covariance.}
 
There are no \redMaGiC\ galaxies in our catalog at redshifts overlapping the
fourth source bin, so the only way to verify the mean redshift of the galaxies
in that bin is to use the COSMOS galaxies. All the other source bins benefit
from the two-fold validation scheme. We therefore checked to see if removing
the highest redshift bin affected our constraints. Figure~\rf{dcut4} shows
that our fiducial constraints are completely consistent with the looser ones
obtained when the highest redshift bin is removed.

 \Sfig{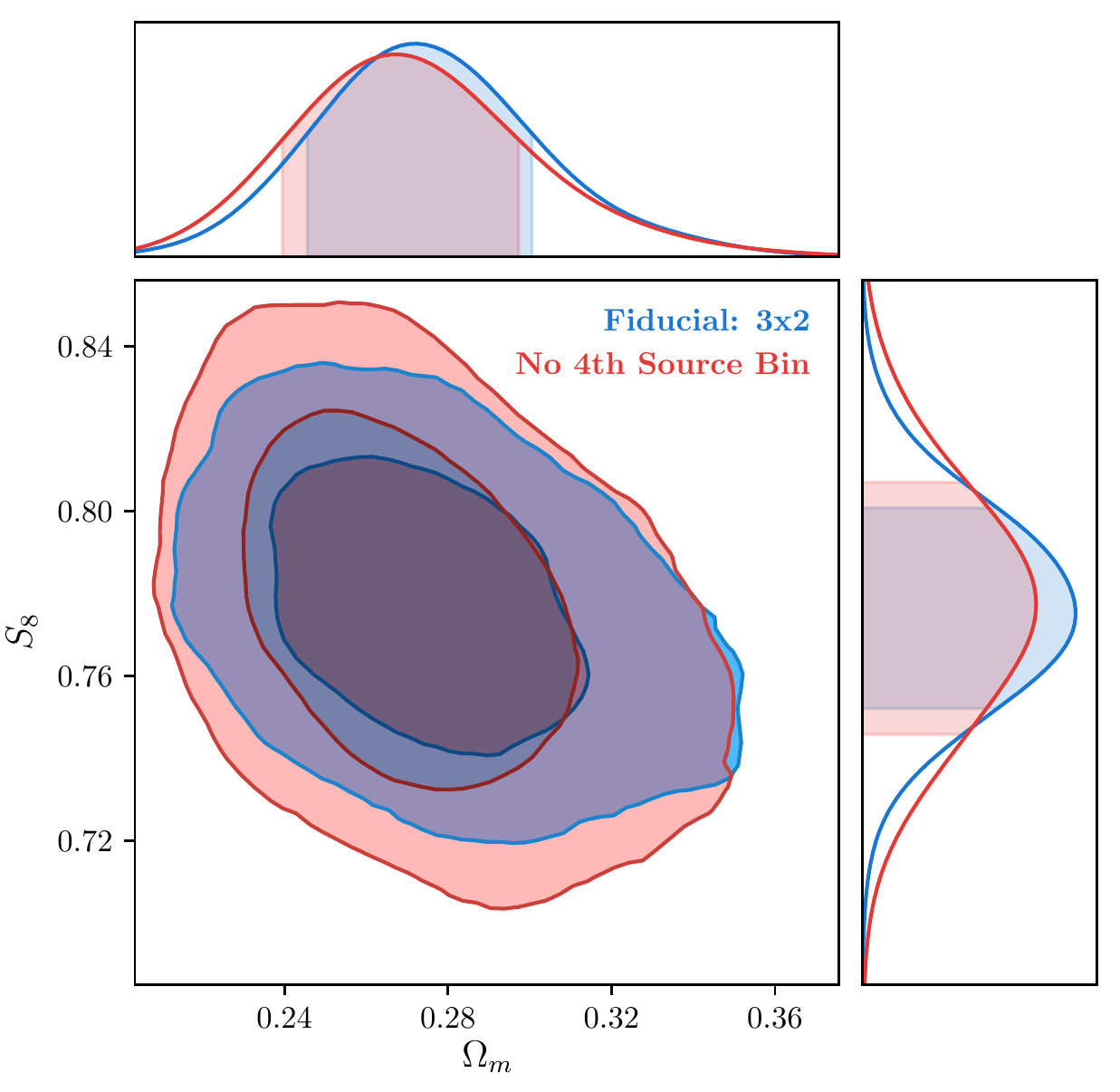}{Constraints from all three probes using all four source bins (``Fiducial'') and with the 4th source bin removed.}

\section{Changes to fiducial covariance}
\label{sec:covupdate}

In the first public version of this paper, the value we reported for
  $\chi^2$ between our fiducial data vector and our best fit model was $\chi^2
  = 572$. This has to be compared to the degrees of freedom of our fit; note
  that $N_{\mathrm{dof}} = N_{\mathrm{data}} - N_{\mathrm{param}}$ is not
  entirely applicable in our situation since we have strong priors on several
  of our parameters. We account for this by assuming an effective number of
  parameters $N_{\mathrm{param,eff.}} = 13$, which is the number of parameters
  that are not tightly constrained by our priors. This resulted in
  $N_{\mathrm{dof, eff}} \approx 457 - 13 = 444$. The reduced $\chi^2$ in the
  first version of the analysis was hence $\approx 1.29$ which, while clearly
  high, was below the threshold of $1.4$ which had been set as a requirement before
  unblinding the analysis.

Following referee comments to the first version of this paper, we were
  fortunately able track down the cause of this elevated $\chi^2$ to two
  inaccuracies of our model covariance:
\begin{enumerate}
\item We had analytically calculated the number of galaxy pairs falling into a
  particular angular bin using simple geometric approximations. These
  approximations can fail for several reasons. First, the finite size of our
  footprint leads to a decrease in the observed density of galaxy pairs found
  on scales comparable to the footprint diameter. Second, the mask pattern on
  scales smaller than the angular scales used in our data vector decreases the
  number of pairs found in each angular bin by a factor that is almost uniform
  across angular scales. And third, clustering of galaxies increases the
  number of galaxy pairs found on small scales. 
\item When estimating the dispersion of intrinsic galaxy shapes
  $\sigma_\epsilon$, we measured the variance of galaxy shape within our
  entire source sample. This ignored propagating the effect of significant differences
  in shape dispersion between different source redshift bins to the covariance
  matrix. 
\end{enumerate}
Both of these analysis improvements affect the noise contribution (shot-noise
and shape-noise) to the diagonal of our covariance matrix. With these changes
identified, we recomputed the shape-noise and shot-noise terms using the
actual numbers of galaxy pairs found in the data and estimating the shape
dispersion separately for each source redshift bin. In addition, we recomputed the
cosmic variance terms in the covariance using our best-fit cosmology.

These changes in our covariance improve the value of $\chi^2$ obtained
  for our best-fit model in the 3x2 $\Lambda$CDM analysis to $497$
  ($\chi^2/{\rm dof} \approx 1.12$). Note that a change in covariance affects
  both the width and the location of parameter constraints, but both of these
  changed very little after the covariance update; by far the dominant effect
  was the improvement in overall $\chi^2$. As an added bonus, the two shear pipelines,
  \metacal\ and \imshape, are now in a better mutual agreement than before
  (see Fig.~\rf{di3}).  

\end{document}

%% file: DES-2017-0226_author_list.tex
% Author list file generated with: mkauthlist 1.2.3 
% mkauthlist -s -j prd DES-2017-0226_author_list.csv DES-2017-0226_author_list.tex 

\author{T.~M.~C.~Abbott}
\affiliation{Cerro Tololo Inter-American Observatory, National Optical Astronomy Observatory, Casilla 603, La Serena, Chile}
\author{F.~B.~Abdalla}
\affiliation{Department of Physics and Electronics, Rhodes University, PO Box 94, Grahamstown, 6140, South Africa}
\affiliation{Department of Physics \& Astronomy, University College London, Gower Street, London, WC1E 6BT, UK}
\author{A.~Alarcon}
\affiliation{Institute of Space Sciences, IEEC-CSIC, Campus UAB, Carrer de Can Magrans, s/n,  08193 Barcelona, Spain}
\author{J.~Aleksi\'c}
\affiliation{Institut de F\'{\i}sica d'Altes Energies (IFAE), The Barcelona Institute of Science and Technology, Campus UAB, 08193 Bellaterra (Barcelona) Spain}
\author{S.~Allam}
\affiliation{Fermi National Accelerator Laboratory, P. O. Box 500, Batavia, IL 60510, USA}
\author{S.~Allen}
\affiliation{Department of Physics, Stanford University, 382 Via Pueblo Mall, Stanford, CA 94305, USA}
\author{A.~Amara}
\affiliation{Department of Physics, ETH Zurich, Wolfgang-Pauli-Strasse 16, CH-8093 Zurich, Switzerland}
\author{J.~Annis}
\affiliation{Fermi National Accelerator Laboratory, P. O. Box 500, Batavia, IL 60510, USA}
\author{J.~Asorey}
\affiliation{School of Mathematics and Physics, University of Queensland,  Brisbane, QLD 4072, Australia}
\affiliation{ARC Centre of Excellence for All-sky Astrophysics (CAASTRO)}
\author{S.~Avila}
\affiliation{Institute of Cosmology \& Gravitation, University of Portsmouth, Portsmouth, PO1 3FX, UK}
\affiliation{Instituto de Fisica Teorica UAM/CSIC, Universidad Autonoma de Madrid, 28049 Madrid, Spain}
% ...

\collaboration{Dark Energy Survey Collaboration}\thanks{For correspondence use des-publication-queries@fnal.gov}

%% file: ack.tex
Funding for the DES Projects has been provided by the U.S. Department of Energy, the U.S. National Science Foundation, the Ministry of Science and Education of Spain, 
the Science and Technology Facilities Council of the United Kingdom, the Higher Education Funding Council for England, the National Center for Supercomputing 
Applications at the University of Illinois at Urbana-Champaign, the Kavli Institute of Cosmological Physics at the University of Chicago, 
the Center for Cosmology and Astro-Particle Physics at the Ohio State University,
the Mitchell Institute for Fundamental Physics and Astronomy at Texas A\&M University, Financiadora de Estudos e Projetos, 
Funda{\c c}{\~a}o Carlos Chagas Filho de Amparo {\`a} Pesquisa do Estado do Rio de Janeiro, Conselho Nacional de Desenvolvimento Cient{\'i}fico e Tecnol{\'o}gico and 
the Minist{\'e}rio da Ci{\^e}ncia, Tecnologia e Inova{\c c}{\~a}o, the Deutsche Forschungsgemeinschaft and the Collaborating Institutions in the Dark Energy Survey. 

The Collaborating Institutions are Argonne National Laboratory, the University of California at Santa Cruz, the University of Cambridge, Centro de Investigaciones Energ{\'e}ticas, 
Medioambientales y Tecnol{\'o}gicas-Madrid, the University of Chicago, University College London, the DES-Brazil Consortium, the University of Edinburgh, 
the Eidgen{\"o}ssische Technische Hochschule (ETH) Z{\"u}rich, 
Fermi National Accelerator Laboratory, the University of Illinois at Urbana-Champaign, the Institut de Ci{\`e}ncies de l'Espai (IEEC/CSIC), 
the Institut de F{\'i}sica d'Altes Energies, Lawrence Berkeley National Laboratory, the Ludwig-Maximilians Universit{\"a}t M{\"u}nchen and the associated Excellence Cluster Universe, 
the University of Michigan, the National Optical Astronomy Observatory, the University of Nottingham, The Ohio State University, the University of Pennsylvania, the University of Portsmouth, 
SLAC National Accelerator Laboratory, Stanford University, the University of Sussex, Texas A\&M University, and the OzDES Membership Consortium.

Based in part on observations at Cerro Tololo Inter-American Observatory, National Optical Astronomy Observatory, which is operated by the Association of 
Universities for Research in Astronomy (AURA) under a cooperative agreement with the National Science Foundation.

The DES data management system is supported by the National Science Foundation under Grant Numbers AST-1138766 and AST-1536171.
The DES participants from Spanish institutions are partially supported by MINECO under grants AYA2015-71825, ESP2015-88861, FPA2015-68048, SEV-2012-0234, SEV-2016-0597, and MDM-2015-0509, 
some of which include ERDF funds from the European Union. IFAE is partially funded by the CERCA program of the Generalitat de Catalunya.
Research leading to these results has received funding from the European Research
Council under the European Union's Seventh Framework Program (FP7/2007-2013) including ERC grant agreements 240672, 291329, and 306478.
We  acknowledge support from the Australian Research Council Centre of Excellence for All-sky Astrophysics (CAASTRO), through project number CE110001020.

This manuscript has been authored by Fermi Research Alliance, LLC under Contract No. DE-AC02-07CH11359 with the 
U.S. Department of Energy, Office of Science, Office of High Energy Physics. 
The United States Government retains and the publisher, by accepting the article for publication, 
acknowledges that the United States Government retains a non-exclusive, paid-up, irrevocable, 
world-wide license to publish or reproduce the published form of this manuscript, or allow 
others to do so, for United States Government purposes.